\documentclass[10pt,twocolumn,twoside]{IEEEtran}
\usepackage{graphicx}
\usepackage{amsmath}
\usepackage{array}
\newcommand{\PreserveBackslash}[1]{\let\temp=\\#1\let\\=\temp}
\newcolumntype{C}[1]{>{\PreserveBackslash\centering}p{#1}}
\newcolumntype{R}[1]{>{\PreserveBackslash\raggedleft}p{#1}}
\newcolumntype{L}[1]{>{\PreserveBackslash\raggedright}p{#1}}
\usepackage[usenames]{color}
\usepackage{colortbl,booktabs}
\usepackage{tabularx}
\usepackage{multicol}
\usepackage{booktabs}
\usepackage[Symbol]{upgreek}
\usepackage{subfigure}
\usepackage{bm}
\usepackage{cite}
\usepackage{balance}
\usepackage[boxed]{algorithm2e}

\usepackage{threeparttable}
\usepackage{amsmath}
\usepackage{dcolumn}
\usepackage{multirow}
\usepackage{booktabs}
\usepackage{amsfonts}
\newcolumntype{d}[1]{D{.}{.}{#1}}% or D{.}{,}{#1} or D{.}{\cdot}{#1}

\def \qed {\hfill \vrule height6pt width 6pt depth 0pt}

\begin{document}

\bibliographystyle{IEEEtran} % use IEEEtran.bst style
\title{Reliable Beamspace Channel Estimation for Millimeter-Wave Massive MIMO Systems with Lens Antenna Array}

\author{Xinyu Gao,~\IEEEmembership{Student Member,~IEEE}, Linglong Dai,~\IEEEmembership{Senior Member,~IEEE}, Shuangfeng Han,~\IEEEmembership{Member,~IEEE},  \\ Chih-Lin I,~\IEEEmembership{Senior Member,~IEEE}, and Xiaodong Wang,~\IEEEmembership{Fellow,~IEEE}

\thanks{Part of this work has been accepted by IEEE/CIC International Conference on Communications in China, Chengdu, China, Jul. 2016.}
\thanks{X. Gao and L. Dai are with the Tsinghua National Laboratory
for Information Science and Technology (TNList), Department of Electronic Engineering, Tsinghua University, Beijing 100084, China (e-mails: xy-gao14@mails.tsinghua.edu.cn, daill@tsinghua.edu.cn).}
\thanks{S. Han and C.-L. I are with the Green Communication Research Center, China Mobile Research
Institute, Beijing 100053, China (e-mails: \{hanshuangfeng, icl\}@chinamobile.com).}
\thanks{X. Wang is with the Electrical Engineering Department, Columbia University,
New York, NY 10027, USA (email: wangx@ee.columbia.edu).}}

%\thanks{This work was supported in part by the International Science \& Technology Cooperation Program of China (Grant No. 2015DFG12760), the National Natural Science Foundation of China (Grant Nos. 61571270 and 61201185),  the Beijing Natural Science Foundation (Grant No. 4142027), and the Foundation of Shenzhen government.}}

\maketitle

\begin{abstract}
Millimeter-wave massive MIMO with lens antenna array can considerably reduce the number of required radio-frequency (RF) chains by beam selection. However, beam selection requires the base station to acquire the accurate information of beamspace channel. This is a challenging task, as the size of beamspace channel is large while the number of RF chains is limited. In this paper, we investigate the beamspace channel estimation problem in mmWave massive MIMO systems with lens antenna array. Specifically, we first design an adaptive selecting network for mmWave massive MIMO systems with lens antenna array, and based on this network, we further formulate the beamspace channel estimation problem as a sparse signal recovery problem. Then, by fully utilizing the structural characteristics of mmWave beamspace channel, we propose a support detection (SD)-based channel estimation scheme with reliable performance and low pilot overhead. Finally, the performance and complexity analyses are provided to prove that the proposed SD-based channel estimation scheme can estimate the support of sparse beamspace channel with comparable or higher accuracy than conventional schemes. Simulation results verify that the proposed SD-based channel estimation scheme outperforms conventional schemes and enjoys satisfying accuracy, even in the low SNR region as the structural characteristics of beamspace channel can be exploited.
\end{abstract}

\begin{keywords}
Massive MIMO, millimeter-wave communications, lens antenna array, beamspace channel estimation.
\end{keywords}

\section{Introduction}\label{S1}

\IEEEPARstart Millimeter-wave (mmWave) massive multiple-input multiple-output (MIMO) has been considered as a key technique for future 5G wireless communications~\cite{han2015large}, since it can achieve significant increase in data rates due to its wider bandwidth~\cite{pi2011introduction} and higher spectral efficiency~\cite{rusek13}. However, realizing mmWave massive MIMO in practice is not a trivial task. One key challenging problem is that each antenna in MIMO systems usually requires one dedicated radio-frequency (RF) chain (including digital-to-analog converter, up converter, etc.)~\cite{wei2014key}. This results in unaffordable hardware cost and energy consumption in mmWave massive MIMO systems, as the number of antennas becomes huge (e.g., 256 antennas)~\cite{han2015large}, and the energy consumption of RF chain is high (e.g., about 250 mW per RF chain at mmWave frequencies, compared with 30 mW per RF chain at cellular frequencies)~\cite{alkhateeb2014mimo}. If we consider the base station (BS) in a typical mmWave massive MIMO system with 256 antennas, only the RF chains will consume 64 Watts, which is much higher than the energy consumption of current 4G micro-cell BS (several Watts)~\cite{han2015large}. To reduce the number of required RF chains, mmWave massive MIMO with lens antenna array has been recently proposed~\cite{brady2013beamspace}.  By employing lens antenna array (an electromagnetic lens with energy focusing capability and a matching antenna array with elements located on the focal surface of the lens), the spatial channel can be transformed into beamspace channel by concentrating the signals from different directions (beams) on different antennas~\cite{brady2013beamspace,zeng16mmwave,Zeng_2014,Behdad10}.
Since the scattering at mmWave frequencies is not rich, the number of effective prorogation paths in mmWave communications is quite limited, occupying only a small number of beams. Therefore,  the mmWave beamspace channel is sparse~\cite{brady2013beamspace}, and we can select a small number of dominant beams according to the sparse beamspace channel to significantly reduce the dimension of MIMO system and the number of required RF chains~\cite{sayeed2013beamspace,amadorilow,gao16bs}. As a result, mmWave massive MIMO with lens antenna array can be considered as a promising solution to relieve the bottleneck of huge energy consumption~\cite{brady2013beamspace}.

Nevertheless, to achieve the capacity-approaching performance, beam selection requires the BS to acquire the information of beamspace channel of large size, which is difficult to realize, especially when the number of RF chains is limited. To solve this problem, some advanced schemes based on compressive sensing (CS) have been proposed very recently~\cite{alkhateeb2014channel,alkhateeb2015compressed,love15,gao2015mmwave,bajwa2010compressed}. The key idea of these schemes is to efficiently utilize the sparsity of mmWave channel in the angle domain. However, these schemes are designed for hybrid precoding systems~\cite{el2013spatially,gao16turbo,gao15energy}, where the phase shifter network is realized by high-resolution phase shifters. As a result, the analog precoders with much higher design freedom can be realized to improve the channel estimation accuracy. By contrast, in mmWave massive MIMO systems with lens antenna array, although the phase shifter network can be replaced by lens antenna array to further reduce the hardware cost and energy consumption, the equivalently designed analog precoders will be restricted to discrete fourier transform (DFT) vectors. From the mathematical view, the analog precoders in mmWave massive MIMO with lens antenna array have stronger constraint~\cite{Behdad10}. As a result, if we directly apply the channel estimation schemes designed for hybrid precoding systems, the performance will be not very satisfying. To estimate the beamspace channel in mmWave massive MIMO systems with lens antenna array, sparsity mask detection (SMD)-based channel estimation is proposed very recently~\cite{Hogan16,Yang16}. The key idea is to first determine which beams with large power should be used (i.e., sparsity mask~\cite{Hogan16,sayeed2013beamspace,song2013beamspace}) by a beam training procedure between the BS and users. Then, we can reduce the dimension of beamspace channel, and estimate the dimension-reduced channel by  classical algorithms, such as least squares (LS). This scheme can efficiently estimate the beamspace channel with quite low computational complexity, and its pilot overhead is also low when the whole channel coherence time or beam coherence time as discussed in~\cite[Section V]{Hogan16} are considered. However, the number of  pilot symbols required to scan all the beams is proportional to the number of BS antennas, which is still large (e.g., 256 antennas).  This can be reduced if we exploit the sparsity of beamspace channel by utilizing CS tools~\cite{bajwa2010compressed}.

In this paper, we propose a support detection (SD)-based channel estimation scheme by utilizing the CS tools to estimate the beamspace channel with low pilot overhead. The contributions of this paper can be summarized as follows:

1) We design an adaptive selecting network, which consists of a small number of 1-bit phase shifters, to replace the recently proposed selecting network in mmWave massive MIMO with lens antenna array~\cite{brady2013beamspace}. For data transmission, the proposed adaptive selecting network can select beams like the traditional one, while for channel estimation, it can perform as a combiner to obtain the efficient measurements of beamspace channel. Then, based on the proposed adaptive selecting network, we formulate the beamspace channel estimation problem as a sparse signal recovery problem~\cite{tropp2006algorithms1}.

2) We propose a SD-based channel estimation scheme. The basic idea is to decompose the total channel estimation problem into a series of sub-problems, each of which only considers one sparse channel component (a vector containing the information of a specific propagation direction)~\cite{gao16iccc}. For each channel component, we first detect its support (i.e., the index set of nonzero elements in a sparse vector) by exploiting the structural characteristics of mmWave beamspace channel. Then, the influence of this channel component is removed, and the support of the next channel component is detected in a similar method. After the supports of all channel components have been detected, the beamspace channel of large size can be estimated with low pilot overhead.

3) We prove that the proposed SD-based channel estimation scheme can detect the supports of channel components more accurately than the classical CS algorithms~\cite{tropp2006algorithms1}. Complexity analysis shows that SD-based channel estimation also enjoys low complexity. Simulation results verify that SD-based channel estimation enjoys satisfying accuracy and low pilot overhead, even in the low signal-to-noise ratio (SNR) region.

The rest of the paper is organized as follows. In Section II, the system model of mmWave massive MIMO with lens antenna array is described. In Section III, we specify the proposed SD-based channel estimation scheme and the analyses of its performance. Finally, the simulation results are provided in Section IV, and conclusions are drawn in Section V.

{\it Notation}: Lower-case and upper-case boldface letters ${\bf{a}}$ and ${\bf{A}}$ denote a vector and a matrix, respectively; ${{{\bf{A}}^H}}$, ${{{\bf{A}}^{ - 1}}}$,  and ${{\rm{tr}}( \bf{A} )}$ denote the conjugate transpose, inversion, and trace of matrix ${\bf{A}}$, respectively; ${{\left\|  \bf{A}  \right\|_F}}$  denotes the Frobenius norm of matrix ${\bf{A}}$; ${{\left\|  \bf{a}  \right\|_2}}$ denotes the ${{l_2}}$-norm of vector ${\bf{a}}$; ${\left|  a  \right|}$ denotes the amplitude of scalar ${a}$; ${{\rm{Card}}\left( {\cal A} \right)}$ denotes the cardinality of set ${{\cal A}}$; Finally, ${{{\mathbf{I}}_{K}}}$ is the ${K \times K}$ identity matrix.

\section{System Model}\label{S2}
We consider a time division duplexing (TDD) mmWave massive MIMO system, where the BS employs ${N}$  antennas and ${{N_{{\rm{RF}}}}}$ RF chains to simultaneously serve ${K}$  single-antenna users~\cite{sayeed2013beamspace,amadorilow,gao16bs}. In this section, we focus on the downlink model to explain the basic principle of mmWave massive MIMO with lens antenna array, while in Section III, the uplink model will be considered for channel estimation, which is just a transposition of the downlink one according to the TDD channel reciprocity.

\begin{figure}[tp]
\begin{center}
\vspace*{0mm}\includegraphics[width=1\linewidth]{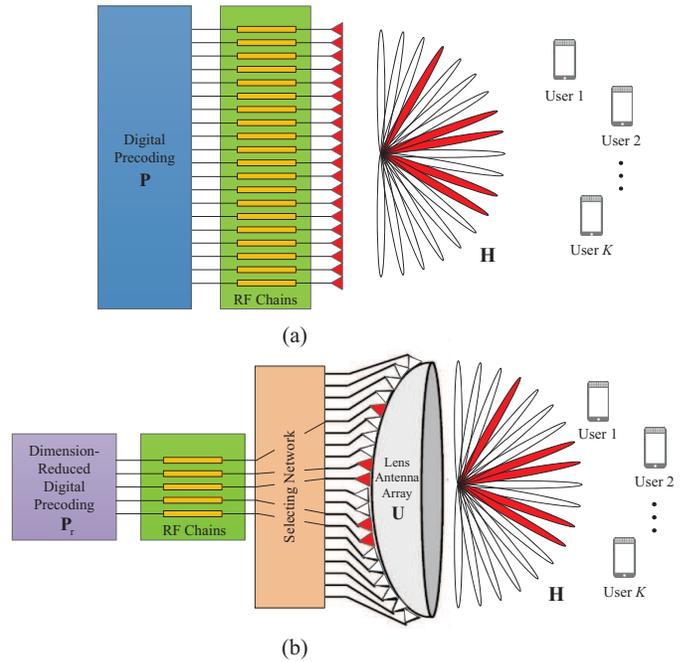}
\end{center}
\vspace*{-3mm}\caption{Comparison of system architectures: (a) traditional mmWave massive MIMO; (b) mmWave massive MIMO with lens antenna array.} \label{FIG1}
\end{figure}

\subsection{Traditional mmWave massive MIMO}\label{S2.1}
Fig. 1 (a) shows the traditional mmWave massive MIMO.  For a narrowband system, the ${K \times 1}$ received signal vector ${{{\bf{y}}^{{\rm{DL}}}}}$ for all ${K}$ users in the downlink can be presented by
\begin{equation}\label{eq1}
{{\mathbf{y}}^{\text{DL}}}={{\mathbf{H}}^{H}}\mathbf{Ps}+\mathbf{n},
\end{equation}
where ${{{\mathbf{H}}^{H}}\in {{\mathbb{C}}^{K\times N}}}$ is the downlink channel matrix, ${\mathbf{H}=\left[ {{\mathbf{h}}_{1}},{{\mathbf{h}}_{2}},\cdots,{{\mathbf{h}}_{K}} \right]}$ is the uplink channel matrix according to the channel reciprocity~\cite{love15}, ${{{\mathbf{h}}_{k}}}$ of size ${N \times 1}$ is the channel vector between the BS and the ${k}$th user as will be discussed in details later, ${{\bf{s}}}$ of size ${K \times 1}$ is the original signal vector for all ${K}$  users with normalized power ${\mathbb{E}\left( \mathbf{s}{{\mathbf{s}}^{H}} \right)={{\mathbf{I}}_{K}}}$, ${{\bf{P}}}$ of size ${N \times K}$ is the precoding matrix satisfying ${{\rm{tr}}\left( {{\bf{P}}{{\bf{P}}^H}} \right) \le \rho }$, where ${\rho }$ is downlink transmit power. Finally, ${\mathbf{n}\sim\mathcal{C}\mathcal{N}\left( 0,\sigma _{\text{DL}}^{2}{{\mathbf{I}}_{K}} \right)}$ is the ${K \times 1}$ additive white Gaussian noise (AWGN) vector, where ${{\sigma _{{\rm{DL}}}^2}}$ is the downlink noise power. It is obvious from Fig. 1 (a) that the number of required RF chains in  traditional MIMO systems is equal to the number of antennas, i.e., ${{N_{{\rm{RF}}}}=N}$, which is usually large for mmWave massive MIMO systems, e.g., ${{N_{{\rm{RF}}}}=N=256}$~\cite{han2015large}.

Next, we will introduce the channel vector ${{{\mathbf{h}}_{k}}}$ of the ${k}$th user. In this paper, we adopt the widely used Saleh-Valenzuela channel model for mmWave communications as~\cite{alkhateeb2014channel}

\begin{equation}\label{eq2}
{{\bf{h}}_k} = \sqrt {\frac{N}{{{L_k} + 1}}} \sum\limits_{i = 0}^{{L_k}} {\beta _k^{\left( i \right)}{\bf{a}}\left( {\psi _k^{\left( i \right)}} \right)}  = \sqrt {\frac{N}{{{L_k} + 1}}} \sum\limits_{i = 0}^{{L_k}} {{{\bf{c}}_{k,i}}},
\end{equation}
where ${{{{\bf{c}}_{k,0}}}=\beta _{k}^{\left( 0 \right)}\mathbf{a}\left( \psi _{k}^{\left( 0 \right)} \right)}$ is the line-of-sight (LoS) component of ${{{\mathbf{h}}_{k}}}$ with ${\beta _k^{\left( 0 \right)}}$ presenting the complex gain and ${{\psi _k^{\left( 0 \right)}}}$ denoting the spatial direction, ${{{{\bf{c}}_{k,i}}}=\beta _{k}^{\left( i \right)}\mathbf{a}\left( \psi _{k}^{\left( i \right)} \right)}$ for ${1\le i\le L_k}$ is the ${i}$th non-line-of-sight (NLoS) component of ${{{\mathbf{h}}_{k}}}$, and ${{{L_k}}}$ is the number of NLoS components, which can be usually obtained by channel measurement~\cite{rappaport2011state}, ${\mathbf{a}\left( \psi  \right)}$ is the ${N \times 1}$ array steering vector. For the typical uniform linear array (ULA) with ${N}$ antennas, we have
\begin{equation}\label{eq3}
{\bf{a}}\left( \psi  \right) = \frac{1}{{\sqrt N }}{\left[ {{e^{ - j2\pi \psi m}}} \right]_{m \in {\cal I}\left( N \right)}},
\end{equation}
where ${{\cal I}\left( N \right) = \left\{ {p - \left( {N - 1} \right)/2,\;p = 0,1, \cdots ,N - 1} \right\}}$ is a symmetric set of indices centered around zero. The spatial direction is defined as ${\psi  \buildrel \Delta \over = \frac{d}{\lambda }\sin \theta }$~\cite{brady2013beamspace}, where ${\theta }$ is the physical direction, ${\lambda }$ is the wavelength of carrier, and ${d}$ is the antenna spacing which usually satisfies ${d = \lambda /2}$ in mmWave communications~\cite{han2015large}.

%\footnote{The extension to 3D scenario with uniform planner array (UPA) is also possible~\cite{brady2014beamspace}.}

\subsection{MmWave massive  MIMO with lens antenna array}\label{S2.2}
The conventional channel (\ref{eq2}) in the spatial domain can be transformed to the beamspace channel by employing a carefully designed lens antenna array~\cite{brady2013beamspace} as shown in Fig. 1 (b). Essentially, such lens antenna array plays the role of a spatial DFT matrix ${{\bf{U}}}$ of size ${N \times N}$, which contains the array steering vectors of ${N}$ orthogonal directions (beams) covering the entire space as
\begin{equation}\label{eq4}
{\bf{U}} = {\left[ {{\bf{a}}\left( {{{\bar \psi }_1}} \right),{\bf{a}}\left( {{{\bar \psi }_2}} \right), \cdots ,{\bf{a}}\left( {{{\bar \psi }_N}} \right)} \right]^H},
\end{equation}
where ${{\bar \psi _n} = \frac{1}{N}\left( {n - \frac{{N + 1}}{2}} \right)}$ for ${n = 1,2, \cdots ,N}$ are the spatial directions pre-defined by lens antenna array. Then, according to  Fig. 1 (b), the system model of mmWave massive MIMO with lens antenna array can be represented by
\begin{equation}\label{eq5}
{{\bf{\tilde y}}^{{\rm{DL}}}} = {{\bf{H}}^H}{{\bf{U}}^H}{\bf{Ps}} + {\bf{n}} = {{\bf{\tilde H}}^H}{\bf{Ps}} + {\bf{n}},
\end{equation}
where ${{{\mathbf{\tilde{y}}}^{\text{DL}}}}$ is the received downlink signal vector in the beamspace, and the beamspace channel ${{{\mathbf{\tilde{H}}}}}$ is defined as
\begin{equation}\label{eq6}
{\bf{\tilde H}}\! =\! \left[ {{{{\bf{\tilde h}}}_1},{{{\bf{\tilde h}}}_2}, \cdots \!,{{{\bf{\tilde h}}}_K}} \right]\! =\! {\bf{UH}} \!=\! \left[ {{\bf{U}}{{\bf{h}}_1},{\bf{U}}{{\bf{h}}_2}, \cdots \!,{\bf{U}}{{\bf{h}}_K}} \right],
\end{equation}
where ${{{{{\bf{\tilde h}}}_k}}}$ is the beamspace channel vector between the BS and the ${k}$th user. It is worth pointing out that the beamspace channel ${{{\mathbf{\tilde{H}}}}}$ (${{{{{\bf{\tilde h}}}_k}}}$) has a sparse structure~\cite{brady2013beamspace,song2013beamspace} due to the limited number of dominant scatterers in the mmWave prorogation environments~\cite{alkhateeb2014channel}. Therefore, as shown in Fig. 2, we can select only a small number of appropriate beams according to the sparse beamspace channel to reduce the dimension of MIMO system as
\begin{equation}\label{eq101}
{{\mathbf{\tilde{y}}}^{\text{DL}}}\approx \mathbf{\tilde{H}}_{\text{r}}^{H}{{\mathbf{P}}_{\text{r}}}\mathbf{s}+\mathbf{n},
\end{equation}
where ${{{\bf{\tilde H}}_{\rm{r}}} = {\bf{\tilde H}}{\left( {b,:} \right)_{b \in {\cal B}}}}$ with ${{\cal B}}$ denoting the sparsity mask (the beam set contains the indices of selected beams)~\cite{Hogan16}, and ${{{\bf{P}}_{\rm{r}}}}$ is the corresponding dimension-reduced digital precoding matrix, which also satisfies the transmit power constraint as ${{\rm{tr}}\left( {{{\bf{P}}_{\rm{r}}}{\bf{P}}_{\rm{r}}^H} \right) \le \rho }$. Due to the sparse nature of the beamspace channel ${{{\mathbf{\tilde{H}}}}}$, mmWave massive MIMO with lens antenna array can significantly reduce the number of required RF chains without obvious performance loss as shown in Fig. 1 (b)~\cite{song2013beamspace,brady2014prototype,brady2016multi}. Note that the smallest number of required RF chains should be ${{N_{{\rm{RF}}}}=K}$ to guarantee the spatial multiplexing gains of ${K}$ users. Therefore, we consider ${{N_{{\rm{RF}}}}=K}$ without loss of generality in this paper.

\begin{figure}[tp]
\begin{center}
\vspace*{0mm}\includegraphics[width=0.7\linewidth]{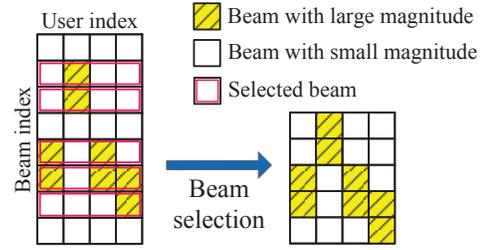}
\end{center}
\vspace*{-3mm}\caption{Illustration of beam selection~\cite{sayeed2013beamspace,amadorilow,gao16bs}.} \label{FIG1}
\end{figure}

To achieve the capacity-approaching performance, beam selection requires the information of beamspace channel ${{\bf{\tilde H}}}$. Fortunately, since the beamforming gains always exist in mmWave massive MIMO systems with lens antenna array, the beamspace channel estimation will not suffer from the serious SNR loss~\cite{Behdad10}. However, as the number of RF chains is limited while the size of beamspace channel is large, estimating the beamspace channel with low pilot overhead is still a challenging problem.

\section{Beamspace Channel Estimation}\label{S3}
In this section, based on the specific architecture of mmWave massive MIMO with lens antenna array, we first introduce a pilot transmission strategy. Then, an adaptive selecting network is designed to obtain the efficient measurements of beamspace channel for channel estimation. After that, a SD-based channel estimation scheme is proposed to estimate the beamspace channel with reliable performance and low pilot overhead. Finally, the performance and complexity analyses are provided to show the advantages of our scheme.

\subsection{Pilot transmission}\label{S3.1}
To estimate the beamspace channel, in the uplink of TDD systems, all users need to transmit the known pilot sequences to the BS over ${Q}$ instants (each user transmits one pilot symbol in each instant) for channel estimation, and we assume that the beamspace channel remains unchanged within such channel coherence time (i.e., ${Q}$ instants)~\cite{tse2005fundamentals}. The impact of channel coherence time on the pilot overhead can be found in~\cite{Kotecha_2004,Hogan16}. It is worth pointing out that for mmWave communications, although the channel coherence time is usually small due to the high carrier frequency, it still contains quite a large number of symbols
thanks to the large mmWave bandwidth~\cite{Hogan16}. For example, when the carrier frequency is 28 GHz and the bandwidth is 1 GHz, a maximum speed of 30 m/s results in the small channel coherence time of 0.36 ms. However, the symbol duration is in the order of 1 ns, which means that the small channel coherence time still contains 400,000 symbols. In this paper, we consider the pilot transmission strategy, where ${Q}$ instants are divided into ${M}$ blocks and each block consists of ${K}$ instants, i.e., ${Q = MK}$\footnote{Note that in practical systems, the number of users may change but the maximum number of users that can be simultaneously served is usually fixed (e.g., at most 4 users in LTE-Advanced systems~\cite{Dongwoon12}). To this end, the total number of pilot symbols ${Q}$ can be designed according to the maximum number. When the number of users exceeds the maximum one, we can resort to user scheduling~\cite{rusek13}.}(the pilot transmission strategy adopted in~\cite{Kotecha_2004,Hogan16} can be also applied here). For the ${m}$th block, we define ${{{\bf{\Psi }}_m}}$ of size ${K \times K}$ as the pilot matrix, which contains ${K}$ mutually orthogonal pilot sequences transmitted by ${K}$ users over ${K}$ instants~\cite{tse2005fundamentals,xie2016unified}.  To normalize the uplink pilot power to unit, we have ${{{\bf{\Psi }}_m}{\bf{\Psi }}_m^H = {{\bf{I}}_K}}$ and ${{\bf{\Psi }}_m^H{{\bf{\Psi }}_m} = {{\bf{I}}_K}}$. A simple example of ${{{\bf{\Psi }}_m}}$ when ${K = 4}$ is
\begin{equation}\label{eq7}
{{\mathbf{\Psi }}_{m}}=\frac{1}{2}\left[ \begin{matrix}
   \text{+}1 & \text{+}1 & \text{+}1 & \text{+}1  \\
   \text{+}1 & \text{+}1 & -1 & -1  \\
   \text{+}1 & -1 & \text{+}1 & -1  \\
   \text{+}1 & -1 & -1 & \text{+}1  \\
\end{matrix} \right].
\end{equation}

Then, according to Fig. 1 (b) and the channel reciprocity~\cite{love15} in TDD systems, the received uplink signal matrix ${\mathbf{\tilde{Y}}_{m}^{\text{UL}}}$ of size ${N \times K}$ at the BS in the ${m}$th block can be presented as
\begin{equation}\label{eq8}
\mathbf{\tilde{Y}}_{m}^{\text{UL}}\!=\!\mathbf{UH}{{\mathbf{\Psi }}_{m}}\!+\!{{\mathbf{N}}_{m}}\!=\!\mathbf{\tilde{H}}{{\mathbf{\Psi }}_{m}}\!+\!{{\mathbf{N}}_{m}},\quad m\!=\!1,2,\cdots,M,
\end{equation}
where ${{{\mathbf{N}}_{m}}}$ is the ${N \times K}$ noise matrix in the ${m}$th block, whose entries are independent and identically distributed (i.i.d.) complex Gaussian random variables with mean zero and variance ${{\sigma _{{\rm{UL}}}^2}}$ (the uplink noise power). As the uplink pilot power is normalized to 1, ${1/\sigma _{{\rm{UL}}}^2}$ can be regarded as the uplink SNR.

%\begin{figure}[tp]
%\begin{center}
%\vspace*{0mm}\includegraphics[width=0.8\linewidth]{pilot1}
%\end{center}
%\vspace*{-3mm}\caption{Illustration of the pilot transmission strategy.} \label{FIG1}
%\end{figure}

\subsection{Adaptive selecting network}\label{S3.2}
We consider the ${m}$th block without loss of generality. During the pilot transmission, the BS should employ a combiner ${{\bf{W}}_m}$ of size ${K \times N}$ to combine the received uplink signal matrix ${\mathbf{\tilde{Y}}_{m}^{\text{UL}}}$~(\ref{eq8}). Then, we can obtain ${{{\bf{R}}_m}}$ of size ${K \times K}$ in the baseband sampled by ${{N_{{\rm{RF}}}} = K}$ RF chains as
\begin{equation}\label{eq13}
{{\mathbf{R}}_{m}}={{\mathbf{W}}_{m}}\mathbf{\tilde{Y}}_{m}^{\text{UL}}={{\mathbf{W}}_{m}}\mathbf{\tilde{H}}{{\mathbf{\Psi }}_{m}}+{{\mathbf{W}}_{m}}{{\mathbf{N}}_{m}}.
\end{equation}
After that, by multiplying the known pilot matrix ${{\bf{\Psi }}_m^H}$ on the right side of~(\ref{eq13}), the ${K \times K}$ measurement matrix ${{{\bf{Z}}_m}}$ of the beamspace channel ${\mathbf{\tilde{H}}}$ can be obtained by
\begin{equation}\label{eq14}
{{\bf{Z}}_m} = {{\bf{R}}_m}{\bf{\Psi }}_m^H = {{\bf{W}}_m}{\bf{\tilde H}} + {\bf{N}}_m^{{\rm{eff}}},
\end{equation}
where ${\mathbf{N}_{m}^{\text{eff}}={{\mathbf{W}}_{m}}{{\mathbf{N}}_{m}}\mathbf{\Psi }_{m}^{H}}$ is the effective noise matrix.

Note that here we focus on estimating the beamspace channel ${{{\bf{\tilde h}}_k}}$ of the ${k}$th user without loss of generality, and the similar method can be directly applied to other users to obtain the complete beamspace channel ${{{\mathbf{\tilde{H}}}}}$. Then, after ${M}$ blocks for the pilot transmission, we can obtain an ${Q \times 1}$ measurement vector ${{{\bf{\bar z}}_k}}$ for ${{{\bf{\tilde h}}_k}}$ as
\begin{equation}\label{eq15}
{{\bf{\bar z}}_k} = \left[ {\begin{array}{*{20}{c}}
{{{\bf{z}}_{1,k}}}\\
{{{\bf{z}}_{2,k}}}\\
 \vdots \\
{{{\bf{z}}_{M,k}}}
\end{array}} \right] = \left[ {\begin{array}{*{20}{c}}
{{{\bf{W}}_1}}\\
{{{\bf{W}}_2}}\\
 \vdots \\
{{{\bf{W}}_M}}
\end{array}} \right]  {{\bf{\tilde h}}_k} + \left[ {\begin{array}{*{20}{c}}
{{\bf{n}}_{1,k}^{{\rm{eff}}}}\\
{{\bf{n}}_{2,k}^{{\rm{eff}}}}\\
 \vdots \\
{{\bf{n}}_{M,k}^{{\rm{eff}}}}
\end{array}} \right] \buildrel \Delta \over = {\bf{\bar W}}{{\bf{\tilde h}}_k} + {{\bf{\bar n}}_k},
\end{equation}
where ${{{\bf{z}}_{m,k}}}$, ${{{\bf{\tilde h}}_k}}$, and ${{\bf{n}}_{m,k}^{{\rm{eff}}}}$ are the ${k}$th column of ${{{\bf{Z}}_m}}$, ${{\bf{\tilde H}}}$, and ${{\bf{N}}_m^{{\rm{eff}}}}$ in~(\ref{eq14}), respectively. ${{{{\bf{\bar z}}}_k}}$, ${{{\bf{\bar W}}}}$, and ${{{{\bf{\bar n}}}_k}}$ are of size ${Q \times 1}$, ${Q \times N}$, and ${Q \times 1}$, respectively. Our target is to reliably reconstruct ${{{\bf{\tilde h}}_k}}$ based on ${{{\bf{\bar z}}_k}}$ with the number of pilot symbols ${Q}$ as low as possible. However, if we directly utilize the recently proposed selecting network~\cite{amadorilow,Hogan16} as shown in Fig. 1 (b) to design ${{{\bf{\bar W}}}}$ (or equivalently ${{\bf{W}}_m}$ for ${m = 1,2, \cdots ,M}$), each row of ${{{\bf{\bar W}}}}$ will have one and only one nonzero element. Consequently, to guarantee that the measurement vector ${{{\bf{\bar z}}_k}}$ contains the complete information of the beamspace channel ${{{\bf{\tilde h}}_k}}$, the number of pilot symbols ${Q}$ should be at least larger than ${N}$, which is still high in mmWave massive MIMO systems as mentioned above.

\begin{figure}[tp]
\begin{center}
\vspace*{0mm}\includegraphics[width=1\linewidth]{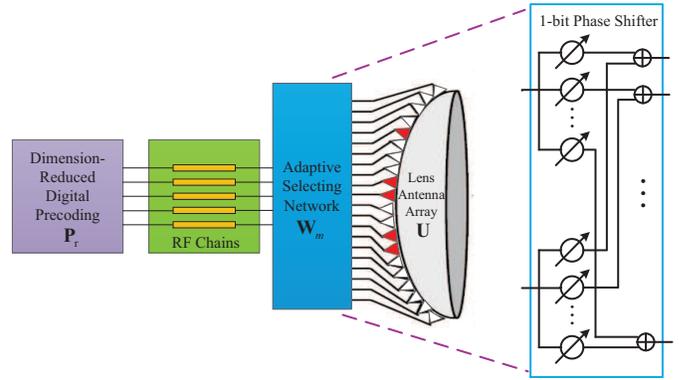}
\end{center}
\vspace*{-3mm}\caption{Proposed adaptive selecting network for mmWave massive MIMO with lens antenna array.} \label{FIG1}
\end{figure}

To this end, we propose an adaptive selecting network for mmWave massive MIMO systems with lens antenna array as shown in Fig. 3, where the selecting network with switches in Fig. 1 (b) is replaced by 1-bit phase shifters. During the data transmission, the proposed adaptive selecting network can be configured to realize the traditional function of beam selection\footnote{Specifically, we can turn off some phase shifters to realize ``unselect"~\cite{Tucker_phase} and set some phase shifters to shift the phase 0 degree to realize ``select" in beam selection.  Alternatively, we can directly utilize the proposed adaptive selecting network to design an analog precoder for data transmission, which can further improve the performance. One possible way is to extend the simple conjugate analog precoder~\cite{Liang_2014} to the scenarios where only 1-bit phase shifters are used. More efficient schemes will be left for our further work.}. Furthermore, during the beamspace channel estimation, this adaptive selecting network can be also adaptively used as an analog combiner ${{\bf{W}}_m}$~\cite{gao15energy} to combine the uplink signals.  Note that ${{{\bf{\tilde h}}_k}}$ is a sparse vector, as the number of dominant scatterers in the mmWave prorogation environments is limited~\cite{sayeed2013beamspace}. Therefore, by utilizing the proposed adaptive selecting network, we can guarantee that ${{{\bf{\bar z}}_k}}$~(\ref{eq15}) has the complete information of ${{{\bf{\tilde h}}_k}}$ even if ${Q < N}$. Then,~(\ref{eq15}) can be formulated as a typical sparse signal recovery problem~\cite{bajwa2010compressed}. The motivation of such formulation is that we can significantly reduce the number of pilot symbols (i.e., ${Q}$ can be much less than ${N}$) at the cost of some SNR loss~\cite{bajwa2010compressed}.

Our next target is to design the analog combiner ${{{\bf{\bar W}}}}$. Under the framework of CS, to achieve the satisfying recovery accuracy, ${{{\bf{\bar W}}}}$ should be designed to make the mutual coherence
\begin{equation}\label{eq103}
\mu  \buildrel \Delta \over = \mathop {\max }\limits_{i \ne j} \left| {{\bf{\bar w}}_i^H{{{\bf{\bar w}}}_j}} \right|
\end{equation}
as small as possible, where ${{{{{\bf{\bar w}}}_i}}}$ is the ${i}$th column of ${{{\bf{\bar W}}}}$. There are already some matrices that have been proved to enjoy small ${\mu}$, such as the i.i.d. Gaussian random matrix and Bernoulli random matrix~\cite{bajwa2010compressed}. In our paper, we select the Bernoulli random matrix as the combiner ${{{\bf{\bar W}}}}$, i.e., each element of ${{{\bf{\bar W}}}}$ is randomly selected from ${\frac{1}{\sqrt{Q}}\left\{ -1,+1 \right\}}$ with equal probability. This is due to the facts that: i) all elements of ${{{\bf{\bar W}}}}$ share the same normalized amplitude, which can be realized by phase shifters; ii) the resolution of phase shifter can be only 1 bit, since we only need to shift the phase by 0 or ${\pi}$. This means that the cost and energy consumption of the phase shifter network can be significantly reduced\footnote{Note that in practice, realizing the phase shifter network with 1-bit phase shifters as shown in Fig. 3 may be also a little complicated. If a certain performance loss is permitted, we can replace the phase shifters by switches, and the proposed scheme can be directly employed without modification.} (1-bit phase shifter is simple and consumes the energy quite similar to that of switch~\cite{mendez2016hybrid}).

\subsection{SD-based channel estimation}\label{S3.3}
After ${{{\bf{\bar W}}}}$ has been designed by the proposed adaptive selecting network,~(\ref{eq15}) can be solved by the classical CS algorithms, such as orthogonal matching pursuit (OMP) and compressive sampling matching pursuit (CoSaMP)~\cite{tropp2007signal}. However, due to the limited transmit power of users, we usually expect that the channel can be estimated with low SNR~\cite{alkhateeb2014channel,zhou2016channel,gao16broadband}. In this case, ${{{\bf{\tilde h}}_k}}$ is overwhelmed by noise, and the support of ${{{\bf{\tilde h}}_k}}$ detected by the classical CS algorithms is inaccurate, leading to the deteriorated performance~\cite{bajwa2010compressed,xie2016overview}. In this paper, by utilizing the structural characteristics of mmWave beamspace channel, we propose a SD-based channel estimation scheme, which can detect the support more accurately and achieve better performance than the classical CS algorithms, especially in the low SNR region. In the following \textbf{Lemma 1}, we will first prove a special property of mmWave beamspace channel, which is one of the two bases of the proposed SD-based channel estimation scheme.

\vspace*{+2mm} \noindent\textbf{Lemma 1}. {\it Represent the beamspace channel ${{{\bf{\tilde h}}_k}}$ as ${{{\bf{\tilde h}}_k} = \sqrt {N/\left( {L_k + 1} \right)} \sum\nolimits_{i = 0}^{{L_k}} {{{{\bf{\tilde c}}}_{k,i}}}}$, where ${{{\bf{\tilde c}}_{k,i}} = {\bf{U}}{{\bf{c}}_{k,i}}}$ is the ${i}$th channel component of ${{{\bf{\tilde h}}_k}}$ in the beamspace. Then, any two channel components ${{{\bf{\tilde c}}_{k,i}}}$ and ${{{\bf{\tilde c}}_{k,j}}}$ in the beamspace are asymptotically orthogonal when the number of antennas ${N}$  in mmWave massive MIMO systems tends to infinity~\cite{Sayeed_2002}, i.e.,}
\begin{equation}\label{eq16}
\mathop {\lim }\limits_{N \to \infty } \left| {{\bf{\tilde c}}_{k,i}^H{{{\bf{\tilde c}}}_{k,j}}} \right| = 0,\quad \forall \;i,j = 0,1, \cdots ,L_k,\quad i \ne j.
\end{equation}

\vspace*{+2mm}
\textit{Proof:} The detailed proof can be found in~\cite{Sayeed_2002}. \qed

\textbf{Lemma 1} implies that for the mmWave massive MIMO system with a large lens antenna array (e.g.,  ${N=256}$), all channel components of ${{{\bf{\tilde h}}_k}}$ in the beamspace are approximately orthogonal to each other. As a result, we can decompose the total channel estimation problem  into a series of independent sub-problems, each of which only considers one specific channel component. Specifically, we can first estimate the strongest channel component. After that, we can remove the influence of this component from the total estimation problem, and then the  channel component with the second strongest power can be estimated. Such procedure will be repeated until all ${\left( {L_k + 1} \right)}$ channel components have been estimated. Next, in the following \textbf{Lemma 2}, we will prove another special structural characteristic of mmWave beamspace channel to show how to estimate each channel component in the beamspace.

\vspace*{+2mm} \noindent\textbf{Lemma 2}. {\it Consider the ${i}$th channel component  ${{{{{\bf{\tilde c}}}_{k,i}}}}$ in the beamspace, and assume ${V}$ is an even integer without loss of generality. The ratio between the power ${{{P_V}}}$ of ${V}$ strongest elements of ${{{{{\bf{\tilde c}}}_{k,i}}}}$ and the total power ${{{P_T}}}$ of ${{{{{\bf{\tilde c}}}_{k,i}}}}$ can be lower-bounded by
\begin{equation}\label{eq20}
\frac{{{P_V}}}{{{P_T}}} \ge \frac{2}{{{N^2}}}\sum\limits_{i = 1}^{V/2} {\frac{1}{{{{\sin }^2}\left( {\frac{{\left( {2i - 1} \right) \pi}}{{2N}}} \right)}}}.
\end{equation}
Moreover, once the position ${n_i^ * }$ of the strongest element of ${{{{{\bf{\tilde c}}}_{k,i}}}}$ is determined, the other ${V-1}$ strongest elements will uniformly locate around it.}

\begin{figure}[tp]
\begin{center}
\vspace*{0mm}\includegraphics[width=1\linewidth]{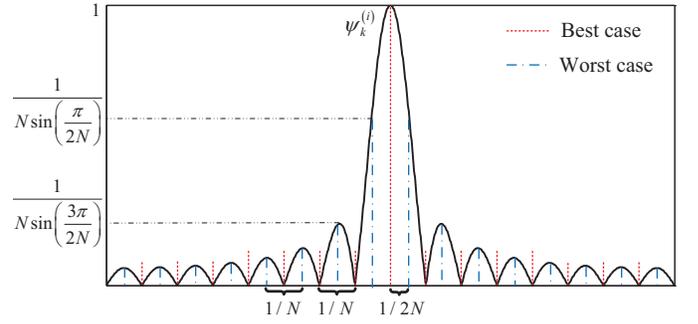}
\end{center}
\vspace*{-3mm}\caption{The normalized amplitude distribution of the elements in ${{{{{\bf{\tilde c}}}_{k,i}}}}$.} \label{FIG1}
\end{figure}

\vspace*{+2mm}
\textit{Proof:} Based on~(\ref{eq2})-(\ref{eq4}), the ${i}$th channel component  ${{{{{\bf{\tilde c}}}_{k,i}}}}$ in the beamspace can be presented as
\begin{equation}\label{eq21}
{{\mathbf{\tilde{c}}}_{k,i}}\!=\!\beta _{k}^{\left( i \right)}{{\left[ \Upsilon \left( {{{\bar{\psi }}}_{1}}\!-\!\psi _{k}^{\left( i \right)} \right),\cdots,\Upsilon \left( {{{\bar{\psi }}}_{N}}\!-\!\psi _{k}^{\left( i \right)} \right) \right]}^{H}},
\end{equation}
where ${\Upsilon \left( x \right) \buildrel \Delta \over = \frac{{\sin N\pi x}}{{N\sin \pi x}}}$. Fig. 4 shows the normalized amplitude distribution of the elements in ${{{{{\bf{\tilde c}}}_{k,i}}}}$, where the set of red dash lines (or blue dot dash lines) presents the set of spatial directions ${{\bar \psi _n} = \frac{1}{N}\left( {n - \frac{{N + 1}}{2}} \right)}$ for ${n = 1,2, \cdots ,N}$ in~(\ref{eq4}). From Fig. 4, we can observe that when the practical spatial direction  ${{\psi _k^{\left( i \right)}}}$ exactly equals one of the pre-defined spatial directions, there is only one strongest element containing all the power of ${{{{{\bf{\tilde c}}}_{k,i}}}}$, which is the best case. In contrast, the worst case will happen when the distance between ${{\psi _k^{\left( i \right)}}}$ and one of the pre-defined spatial directions is equal to ${1/2N}$. In this case, the power ${{{P_V}}}$ of ${V}$ strongest elements of ${{{{{\bf{\tilde c}}}_{k,i}}}}$ is
\begin{equation}\label{eq22}
{P_V} = \frac{{2{{\left( {\beta _k^{\left( i \right)}} \right)}^2}}}{{{N^2}}}\sum\limits_{i = 1}^{V/2} {\frac{1}{{{{\sin }^2}\left( {\frac{{\left( {2i - 1} \right)\pi }}{{2N}}} \right)}}}.
\end{equation}

Besides, according to~(\ref{eq21}), the total power ${{P_T}}$ of ${{{{{\bf{\tilde c}}}_{k,i}}}}$ can be calculated as
\begin{equation}\label{eq101}
{P_T} = {\bf{\tilde c}}_{k,i}^H{{\bf{\tilde c}}_{k,i}} = {\left( {\beta _k^{\left( i \right)}} \right)^2}.
\end{equation}
Based on~(\ref{eq22}) and~(\ref{eq101}), we can conclude that ${{P_V}/{P_T}}$ is lower-bounded by
\begin{equation}\label{eq23}
\frac{{{P_V}}}{{{P_T}}} \ge \frac{2}{{{N^2}}}\sum\limits_{i = 1}^{V/2} {\frac{1}{{{{\sin }^2}\left( {\frac{{\left( {2i - 1} \right) \pi}}{{2N}}} \right)}}}.
\end{equation}
Moreover, as shown in Fig. 4, once the position ${n_i^ * }$ of the strongest element of ${{{{{\bf{\tilde c}}}_{k,i}}}}$ is determined, the other ${V-1}$ strongest elements will uniformly locate around it. \qed

\begin{algorithm}[tp]
\caption{Proposed SD-based channel estimation.}
\KwIn{\\
\\\hspace*{+2mm} Measurement vector: ${{{\bf{\bar z}}_k}}$ in~(\ref{eq15});
\\\hspace*{+2mm} Combining matrix: ${{{\bf{\bar W}}}}$ in~(\ref{eq15});
\\\hspace*{+2mm} Total number of channel components: ${L_k + 1}$;
\\\hspace*{+2mm} Retained number of elements for each component: ${V}$.}
\textbf{Initialization}: ${{\bf{\tilde c}}_{k,i}^{\rm{e}} = {{\bf{0}}_{N \times 1}}}$ for ${0 \le i \le L_k}$, ${{\bf{\bar z}}_k^{\left( 0 \right)} = {{\bf{\bar z}}_k}}$.
\\\textbf{for} ${0 \le i \le L_k}$
 \\1. Detect the position of the strongest element of ${{{{{\bf{\tilde c}}}_{k,i}}}}$ as
 \\\hspace*{+4mm} ${n_i^ *  = \mathop {\arg \max }\limits_{1 \le n \le N} \left| {{\bf{\bar w}}_n^H{\bf{\bar z}}_k^{\left( i \right)}} \right|}$, ${{{{{\bf{\bar w}}}_n}}}$ is the ${n}$th column of ${{{\bf{\bar W}}}}$;
 \\2. Detect ${{\rm{supp}}\left( {{{{\bf{\tilde c}}}_{k,i}}} \right)}$ according to~(\ref{eq24});
 \\3. LS estimation of the nonzero elements of ${{{{{\bf{\tilde c}}}_{k,i}}}}$ as
 \\\hspace*{+2mm} ${{{\bf{f}}_i} = {\left( {{\bf{\bar W}}_i^H{{{\bf{\bar W}}}_i}} \right)^{ - 1}}{\bf{\bar W}}_i^H{\bf{\bar z}}_k^{\left( i \right)}}$, ${{{\bf{\bar W}}_i} = {\bf{\bar W}}{\left( {:,b} \right)_{b \in {\rm{supp}}\left( {{{{\bf{\tilde c}}}_{k,i}}} \right)}}}$;
 \\4. Form the estimated ${{\bf{\tilde c}}_{k,i}^{\rm{e}}}$ as ${{\bf{\tilde c}}_{k,i}^{\rm{e}}\left( {{\rm{supp}}\left( {{{{\bf{\tilde c}}}_{k,i}}} \right)} \right) = {{\bf{f}}_i}}$;
 \\5. Remove the influence of ${{{{{\bf{\tilde c}}}_{k,i}}}}$ by ${{\bf{\bar z}}_k^{\left( {i + 1} \right)} = {\bf{\bar z}}_k^{\left( i \right)} - {\bf{\bar W\tilde c}}_{k,i}^{\rm{e}}}$
 \\6. ${i=i+1}$;
\\\textbf{end for}
\\7. ${{{\mathcal{S}}_{\text{T}}}=\underset{0\le i\le L_k}{\mathop{\bigcup }}\,\text{supp}\left( {{{\mathbf{\tilde{c}}}}_{i}} \right)}$;
\\8. ${{{\bf{f}}_{\rm{T}}} = {\left( {{\bf{\bar W}}_{\rm{T}}^H{{{\bf{\bar W}}}_{\rm{T}}}} \right)^{ - 1}}{\bf{\bar W}}_{\rm{T}}^H{{\bf{\bar z}}_k}}$, ${{{\bf{\bar W}}_{\rm{T}}} = {\bf{\bar W}}{\left( {:,b} \right)_{b \in {{\cal S}_{\rm{T}}}}}}$;
\\9. ${{\bf{\tilde h}}_k^{\rm{e}} = {{\bf{0}}_{N \times 1}}}$, ${{\bf{\tilde h}}_k^{\rm{e}}\left( {{{\cal S}_{\rm{T}}}} \right) = {{\bf{f}}_{\rm{T}}}}$;
\\\KwOut{Estimated beamspace channel for user ${k}$: ${{\bf{\tilde h}}_k^{\rm{e}}}$.}
\end{algorithm}

Note that by considering the channel with only LoS component,~\cite{sayeed2013beamspace} has analyzed the power ratio between the total channel and the approximated channel with only two strongest beams. The method used in~\cite[Appendix]{sayeed2013beamspace} can be also modified and extended to prove \textbf{Lemma 2} in this paper. From \textbf{Lemma 2}, we can derive two important conclusions. The first one is that ${{{{{\bf{\tilde c}}}_{k,i}}}}$  can be considered as a sparse vector, since the most power of ${{{{{\bf{\tilde c}}}_{k,i}}}}$  is focused on a small number of dominant elements. For example, when ${N=256}$ and ${V=8}$, the lower-bound of ${{P_V}/{P_T}}$ is about 95\%. This means that we can retain only a small number (e.g., ${V=8}$) of elements of ${{{{{\bf{\tilde c}}}_{k,i}}}}$ with strong power, and regard other elements as zero without obvious performance loss. The second one is that the support of the sparse channel component ${{{{\bf{\tilde c}}}_{k,i}}}$ can be uniquely determined by ${n_i^ * }$ as\footnote{Correspondingly, when ${V}$ is odd, the support of ${{{{{\bf{\tilde c}}}_{k,i}}}}$ should be ${{\rm{supp}}\left( {{{{\bf{\tilde c}}}_{k,i}}} \right) = \,\bmod \,{{\mkern 1mu} _N}\left\{ {n_i^ *  - \frac{{V - 1}}{2}, \cdots ,n_i^ *  + \frac{{V - 1}}{2}} \right\}}$.}
\begin{equation}\label{eq24}
{\rm{supp}}\left( {{{{\bf{\tilde c}}}_{k,i}}} \right) = \,\bmod \,{{\mkern 1mu} _N}\left\{ {n_i^ *  - \frac{V}{2}, \cdots ,n_i^ *  + \frac{{V - 2}}{2}} \right\},
\end{equation}
where ${\text{Card}\left( \text{supp}\left( {{{\mathbf{\tilde{c}}}}_{k,i}} \right) \right)=V}$, and ${{\bmod _N}\left(  \cdot  \right)}$ is the modulo operation with respect to ${N}$, which guarantees that all indices in ${{\rm{supp}}\left( {{{{\bf{\tilde c}}}_{k,i}}} \right)}$ belong to ${\left\{ {1,2, \cdots ,N} \right\}}$. After the support of ${{{{{\bf{\tilde c}}}_{k,i}}}}$ has been detected, we can extract ${V}$ columns from ${{\bf{\bar W}}}$~(\ref{eq15}) according to ${{\rm{supp}}\left( {{{{\bf{\tilde c}}}_{k,i}}} \right)}$, and use the classical LS algorithm to estimate the nonzero elements of ${{{{{\bf{\tilde c}}}_{k,i}}}}$.

%\footnote{In general, the total number of channel components ${\left( {L + 1} \right)}$ can be obtained by channel measurement~\cite{rappaport2011state}.}

Based on the discussion above, the pseudo-code of the proposed SD-based channel estimation can be summarized in \textbf{Algorithm 1}, which can be explained as follows. During the ${i}$th iteration, we first detect the position ${n_i^ * }$ of the strongest element of ${{{{{\bf{\tilde c}}}_{k,i}}}}$ in step 1 by utilizing the low mutual coherence property of ${{\bf{\bar W}}}$ (\ref{eq15}), which is the same as the classical CS algorithms~\cite{tropp2007signal}. Then in step 2, utilizing the structural characteristic of mmWave beamspace channel as analyzed before, we can directly obtain ${{{\rm{supp}}\left( {{{{\bf{\tilde c}}}_{k,i}}} \right)}}$ according to~(\ref{eq24}), which is the key step of the proposed SD-based channel estimation scheme as will be discussed later. After that, the nonzero elements of ${{{{{\bf{\tilde c}}}_{k,i}}}}$ are estimated by LS algorithm in step 3, and the influence of this channel component  is removed in steps 4 and 5. Such procedure will be repeated (${i=i+1}$ in step 6) until all ${\left( {L_k + 1} \right)}$ channel components have been considered.
It is worth pointing out that for the proposed SD-based channel estimation scheme, we do not directly estimate the beamspace channel as ${{\bf{\tilde h}}_k^{\rm{e}} = \sqrt {\frac{N}{{L_k + 1}}} \sum\limits_{i = 0}^{L_k} {{\bf{\tilde c}}_{k,i}^{\rm{e}}}}$. This is because that most of the elements with small power are regarded as zero, which will lead to error propagation in the influence removal (step 5), especially when ${i}$ is large. As a result, ${{\bf{\bar z}}_k^{\left( {i} \right)}}$ will be more and more inaccurate to estimate the nonzero elements in step 3. To this end, we only utilize ${{\bf{\bar z}}_k^{\left( {i} \right)}}$ to estimate the position ${{n_i^ * }}$ of the strongest element of ${{{{{\bf{\tilde c}}}_{k,i}}}}$  in step 1, which can still guarantee a satisfying recovery probability even if ${{\bf{\bar z}}_k^{\left( {i} \right)}}$ is a little inaccurate. Then, after the iterative procedure, we can obtain the total support ${{{\cal S}_{\rm{T}}}}$ of ${{{\bf{\tilde h}}_k}}$  in step 7\footnote{In the proposed SD-based channel estimation scheme, we estimate ${{\rm{Card}}\left( {{{\cal S}_{\rm{T}}}} \right)}$ nonzero elements for ${{{\bf{\tilde h}}_k}}$. Although some of these nonzero elements may be deleted after beam selection, they are still useful to determine the set of selected beams to eliminate the multi-user interferences, just like the beam selection schemes proposed in~\cite{gao16bs,amadorilow}. Moreover, it is worth pointing out that estimating a little more nonzero elements does not require more pilot symbols. The only cost is the slightly increased computational complexity, as will be analyzed in Section III-E.}. Using ${{{\cal S}_{\rm{T}}}}$ and ${{{\bf{\bar z}}_k}}$, we can alleviate the impact of error propagation and estimate the beamspace channel more accurately in steps 8 and 9.

The key difference between \textbf{Algorithm 1} and the classical CS algorithms~\cite{tropp2007signal} is the support detection in  step 2. In the classical CS algorithms, all the positions of nonzero elements are estimated one by one in an iterative procedure, which may be inaccurate, especially for the element whose power is not strong enough. By contrast, in our algorithm, we only estimate the position of the strongest element. Then, by utilizing the structural characteristics of mmWave beamspace channel, we can directly obtain the support with higher accuracy. These conclusions analyzed above will be proved in the following subsection.

\subsection{Performance analysis of SD-based channel estimation}\label{S3.4}
In this subsection, we will prove that the proposed SD-based channel estimation can detect the support more accurately than the classical CS algorithms.

From \textbf{Algorithm 1}, we can observe that the accuracy of the detected support is determined by step 1, i.e., the estimation of the position of the strongest element. Therefore, to evaluate the accuracy of the detected support, we need to analyze the probability of the event ``the position of the strongest element is correctly estimated" as shown in the following \textbf{Lemma 3}.

\vspace*{+2mm} \noindent\textbf{Lemma 3}. {\it Consider the LoS scenario, i.e., ${{{\mathbf{\tilde{h}}}_{k}}=\sqrt{N}{{\mathbf{\tilde{c}}}_{k,0}}}$ and suppose that the strongest element ${{{\tilde{h}}_{k,{{n}^{*}}}}}$ of ${{{\mathbf{\tilde{h}}}_{k}}}$ satisfies\footnote{Here we only consider the LoS scenario for the expression simplicity. It is worth pointing out that the conclusions in \textbf{Lemma 3} can be also extended to the scenario with NLoS components, where the only changes are the expressions of ${\eta }$~(\ref{eq26}) and ${\kappa }$~(\ref{eq27}) as verified in Appendix A.} }
\begin{equation}\label{eq25}
\left| {{{\tilde{h}}}_{k,{{n}^{*}}}} \right|\ge \frac{\sqrt{8\sigma _{\text{UL}}^{2}\left( 1+\alpha  \right)\ln N}}{\left( 1-\mu  \right)\left( 1-\kappa  \right)-2\mu \eta },
\end{equation}
{\it where ${\alpha >0}$ is a constant, and we define that}
\begin{equation}\label{eq26}
\eta  \buildrel \Delta \over = \frac{{\sum\limits_{n = 1}^N {\left| {\frac{1}{{\sin \left( {\left( {2n - 1} \right)\pi /2N} \right)}}} \right|}  - \left| {\frac{1}{{\sin \left( {\pi /2N} \right)}}} \right|}}{{\left| {\frac{1}{{\sin \left( {\pi /2N} \right)}}} \right|}},
\end{equation}
\begin{equation}\label{eq27}
\kappa  \buildrel \Delta \over = \left| {\frac{{\sin \left( {\pi /2N} \right)}}{{\sin \left( {3\pi /2N} \right)}}} \right|.
\end{equation}
{\it Then, the probability ${\Pr 1}$ that the position of the strongest element is correctly estimated is lower-bounded by}
\begin{equation}\label{eq28}
\Pr 1 \ge {\left( {1 - \frac{1}{{{N^{\alpha  + 1}}\sqrt {\pi \left( {1 + \alpha } \right)\ln N} }}} \right)^N}.
\end{equation}

\vspace*{+2mm}
\textit{Proof:} See Appendix A. \qed

\begin{figure}[tp]
\begin{center}
\vspace*{0mm}\includegraphics[width=1\linewidth]{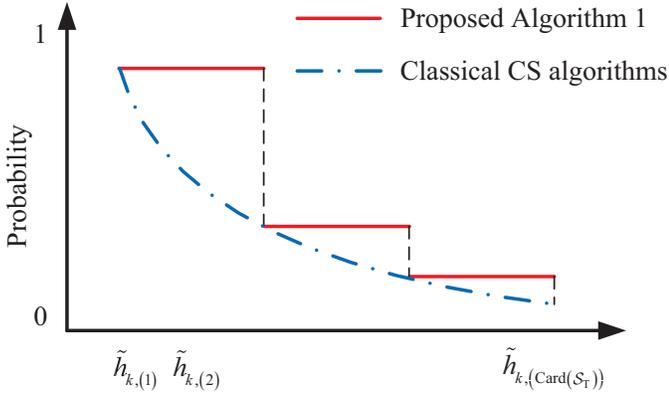}
\end{center}
\vspace*{-3mm}\caption{An illustration of the probability comparison.} \label{FIG3}
\end{figure}

In~(\ref{eq25}), ${\eta }$ and ${\kappa }$ are constants, and ${\mu }$ is fixed when the combiner ${{\bf{\bar W}}}$ has been designed. Therefore, the constant ${\alpha }$ can be interpreted as the effective SNR in general. Given the amplitude of the strongest element ${\left| {{{\tilde h}_{{k, n^ * }}}} \right|}$, for the large uplink noise power ${\sigma _{{\rm{UL}}}^2}$, the constant ${\alpha }$ should be small. Consequently, it will be more difficult to correctly estimate the position of the strongest element according to~(\ref{eq28}). In addition, for a large number of pilot symbols ${Q}$, ${\mu }$ in~(\ref{eq25}) is more likely to be small~\cite[Lemma 8]{tropp2007signal}. As a result, ${\alpha}$ should be large, leading to a higher probability to correctly estimate the position of the strongest element.

From \textbf{Lemma 3}, we can conclude that the proposed SD-based channel estimation scheme can obtain the support with higher accuracy than the classical CS algorithms, which can be explained as follows: For SD-based channel estimation, once the position of the strongest element ${{{{\tilde h}_{{k,n^ * }}}}}$ has been estimated, the positions of the rest of elements can be directly obtained from~(\ref{eq24}), which means that the probability to correctly estimate the positions of all nonzero elements can be lower-bounded by~(\ref{eq28}). In contrast, for the classical CS algorithms, the positions of all elements are estimated one by one in an iterative procedure. Take the second strongest element ${{{{\tilde h}_{{k, n^{ *  * }}}}}}$ for example. In this case, ${\left| {{{\tilde h}_{{k, n^ * }}}} \right|}$ in~(\ref{eq25}) should be replaced by ${\left| {{{\tilde h}_{{k, n^{ *  * }}}}} \right|}$ (smaller than ${\left| {{{\tilde h}_{{n^ * }}}} \right|}$), and the estimate error of the strongest element will equivalently lead to the increase of ${\sigma _{{\rm{UL}}}^2}$. As a result,  ${\alpha }$ will be small and the probability ${\Pr 1}$ will decrease. This indicates that for the classical CS algorithms, the probability to correctly estimate the position of one element drops as the amplitude of this element decreases.

Such conclusion can be further  intuitively illustrated by Fig. 5, where ${{\tilde h_{k,\left( n \right)}}}$ is the ${n}$th strongest nonzero element of ${{{\bf{\tilde h}}_k}}$ (i.e., ${\left| {{{\tilde h}_{k,\left( 1 \right)}}} \right| \ge \left| {{{\tilde h}_{k,\left( 2 \right)}}} \right| \ge  \cdots  \ge \left| {{{\tilde h}_{k,\left( {{\rm{Card}}\left( {{{\cal S}_{\rm{T}}}} \right)} \right)}}} \right|}$). The curve of \textbf{Algorithm 1} in Fig. 5 is divided into several segments due to the fact that we consider each channel component independently. From Fig. 5, we can observe that our algorithm is expected to detect the support more accurately, which will be also verified by simulation results in Section IV.

\subsection{Complexity analysis of SD-based channel estimation}\label{S3.5}
In this section, the complexity of SD-based channel estimation is discussed.

From \textbf{Algorithm 1}, we can observe that the complexity of the proposed SD-based channel estimation scheme mainly comes from steps 1, 3, 5, and 8. In step 1, we need to compute ${N}$ inner products between two ${Q \times 1}$ vectors ${{{\bf{\bar w}}_n}}$ and ${{{\bf{\bar z}}_k^{\left( i \right)}}}$, whose complexity is ${{\cal O}\left( {QN} \right)}$. In step 3, since ${{{\bf{\bar W}}_i}}$ is of size ${Q \times V}$, the LS algorithm can be employed with the complexity ${{\cal O}\left( {{V^2}Q} \right)}$. In step 5, the influence of the ${i}$th channel component is removed by calculating the multiplication between the ${Q \times N}$ matrix ${{\bf{\bar W}}}$ and the ${N \times 1}$ vector ${{\bf{\tilde c}}_{k,i}^{\rm{e}}}$, which has the complexity ${{\cal O}\left( {QN} \right)}$. Finally, in step 8, LS algorithm is used again with the complexity ${{\cal O}\left( {{\rm{Car}}{{\rm{d}}^2}\left( {{{\cal S}_{\rm{T}}}} \right)Q} \right)}$ as ${{{\bf{\bar W}}_{\rm{T}}}}$ is of size ${Q \times {\rm{Card}}\left( {{{\cal S}_{\rm{T}}}} \right)}$.

To sum up, the  complexity of the proposed SD-based channel estimation scheme can be presented as
\begin{equation}\label{eq29}
{\cal O}\left( {{L_k}{V^2}Q} \right) + {\cal O}\left( {{L_k}NQ} \right) + {\cal O}\left( {{\rm{Car}}{{\rm{d}}^2}\left( {{S_{\rm{T}}}} \right)Q} \right),
\end{equation}
where we have ${\text{Card}\left( {{\mathcal{S}}_{\text{T}}} \right)\le V{L_k}}$ according to step 7 in \textbf{Algorithm 1}. Since ${L_k}$ and ${V}$ are usually small as we have discussed before, we can conclude that the  complexity of SD-based channel estimation is quite low, as it is comparable with that of LS algorithm. Finally, it should be also pointed out that the SMD-based channel estimation proposed in~\cite{Hogan16} enjoys much lower complexity compared with our scheme, as the dimension-reduced beamspace channel can be directly obtained by scanning all the beams. In addition, it also enjoys low pilot overhead when the whole channel/beam coherence time is considered, since the symbol duration is quite small due to the large bandwidth and there are a large number of symbols in the channel/beam coherence time~\cite[Table II]{Hogan16}. However, to scan all the beams, the number of required pilot symbols in the SMD-based channel estimation scheme is still proportional to ${N}$, which is a large number (e.g., ${N = 256}$). By contrast, our scheme can further reduce the number of pilot symbols by utilizing the CS tools as will be shown in Section IV.

\section{Simulation Results}\label{S5}
In this section, we consider a typical mmWave massive MIMO system, where the BS equips a lens antenna array with ${N=256}$ antennas and ${{N_{{\rm{RF}}}}{\rm{ = 16}}}$ RF chains to serve ${K=16}$ users. For the ${k}$th user, the spatial channel is generated as follows~\cite{sayeed2013beamspace}: 1) one LoS component and ${L_k=2}$ NLoS components; 2) ${\beta _k^{\left( 0 \right)} \sim {\cal C}{\cal N}\left( {0,1} \right)}$, and ${\beta _k^{\left( i \right)} \sim {\cal C}{\cal N}\left( {0,{{10}^{ - 0.5}}} \right)}$ for ${i = 1,2}$; 3) ${\psi _{k}^{\left( 0 \right)}}$ and ${\psi _{k}^{\left( i \right)}}$  follow the i.i.d. uniform distribution within ${\left[ -0.5,0.5 \right]}$. Finally, the uplink and downlink SNR are defined as ${1/\sigma _{{\rm{UL}}}^2}$ and ${{\rho}/\sigma _{{\rm{DL}}}^2}$, respectively.

Fig. 6 shows the normalized mean square error (NMSE) performance comparison among the SMD-based~\cite{Hogan16}, OMP-based~\cite{mendez2015channel}, and the SD-based channel estimation schemes. For SD-based channel estimation, we retain ${V = 8}$ strongest elements as analyzed above for each channel component. For OMP-based channel estimation~\cite{mendez2015channel}, we assume that the sparsity level of the beamspace channel is equal to ${V\left( L_k + 1 \right)=24}$, and we also make a small modification on the designed sensing matrix to enable it to be applied in mmWave massive MIMO systems with lens antenna array. Note that there are at most ${V(L_k+1)}$ nonzero elements of ${{{\bf{\tilde h}}_k}}$ to be estimated. Therefore, the total number of instants ${Q}$ for pilot transmission  should be at least lager than ${V\left( {\mathop {\max }\limits_k \left( {{L_k}} \right) + 1} \right) = 24}$. In this section, we assume that both SD-based and OMP-based channel estimation schemes employ ${Q = 96}$ instants (i.e., ${M = 6}$ blocks), and the SMD-based channel estimation scheme employ ${Q = N = 256}$ instants for pilot transmission.

\begin{figure}[tp]
\begin{center}
\vspace*{0mm}\includegraphics[width=1\linewidth]{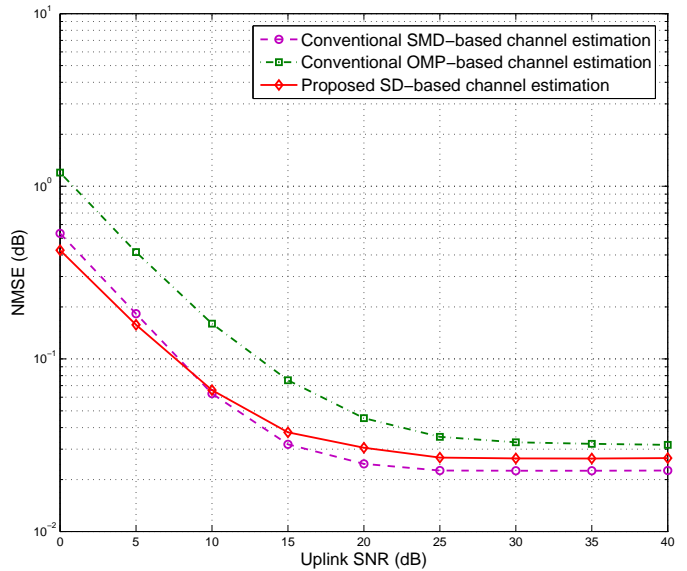}
\end{center}
\vspace*{-3mm}\caption{NMSE performance comparison among SMD-based, OMP-based, and SD-based channel estimation.} \label{FIG3}
\end{figure}

\begin{figure}[tp]
\begin{center}
\vspace*{+1mm}\includegraphics[width=1\linewidth]{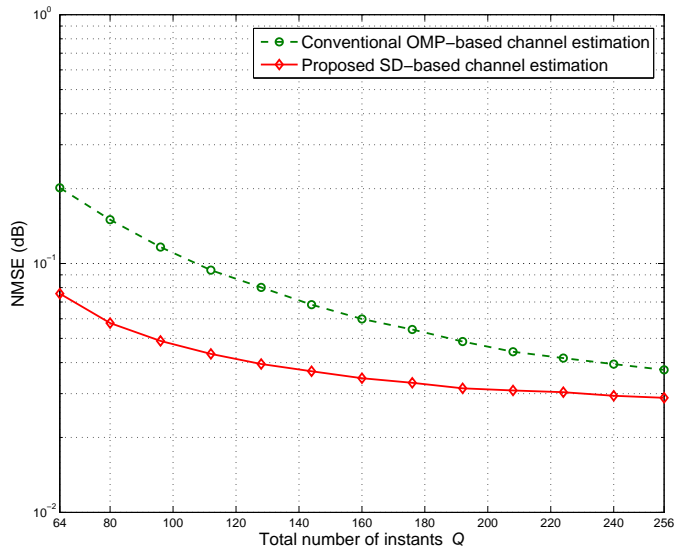}
\end{center}
\vspace*{-3mm}\caption{NMSE performance comparison against the total number of instants ${Q}$ for pilot transmission.} \label{FIG3}
\end{figure}

From Fig. 6, we first observe that the proposed SD-based channel estimation with ${Q = 96}$ instants can achieve the NMSE performance close to the SMD-based channel estimation in [21] with ${Q = N = 256}$ instants. In addition, we also observe that SD-based channel estimation enjoys higher accuracy than OMP-based channel estimation when the uplink SNR is low (e.g., less than 15 dB). When the uplink SNR is high, the performance gap becomes smaller. This can be explained by the fact that when the uplink SNR is high and the total number of instants ${Q}$ is large enough, OMP-based channel estimation can also accurately estimate the support of the beamspace channel. However, the uplink SNR is usually low due to the limited transmit power of users~\cite{alkhateeb2014channel}. Therefore, we can conclude that the proposed SD-based channel estimation scheme is attractive. Finally, Fig. 6 shows that when the uplink SNR is high enough, the NMSE performance of all considered channel estimation schemes will saturate. This is due to the fact that although the nonzero elements of beamspace channel can be estimated accurately with sufficiently high uplink SNR, the error induced by regarding the elements with small power as zero does not vanish.

Fig. 7 shows the NMSE performance comparison against the total number of instants ${Q}$, where the uplink SNR is set as 10 dB. From Fig. 7, we can observe that to achieve the same accuracy, the total number of instants ${Q}$ required by SD-based channel estimation is much lower than OMP-based channel estimation. For example, to achieve the NMSE of
${5 \times {10^{ - 2}}}$, the total number of instants required by OMP-based channel estimation is ${Q = 190}$, while the proposed SD-based channel estimation scheme only requires ${Q = 120}$ instants. Besides, since the number of instants ${Q}$ required in SMD-based channel estimation is larger than ${N}$ (e.g., ${Q \ge N = 256}$), we can also conclude that the proposed SD-based channel estimation scheme can achieve satisfying performance with further reduced pilot overhead.

\begin{figure}[tp]
\begin{center}
\vspace*{0mm}\includegraphics[width=1\linewidth]{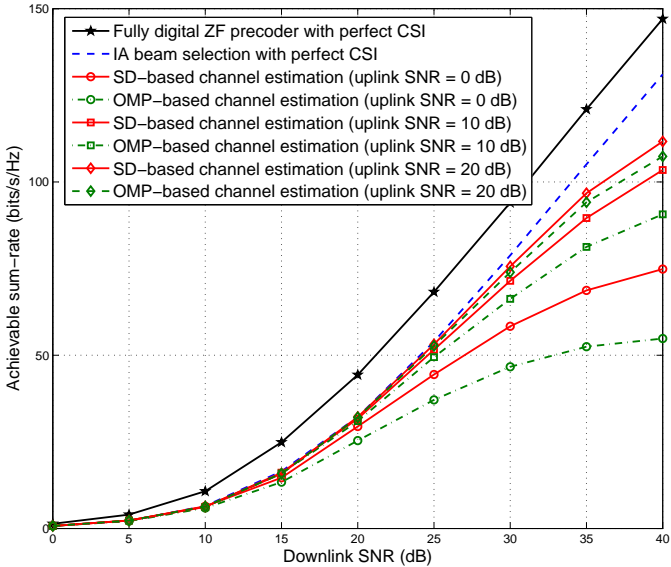}
\end{center}
\vspace*{-3mm}\caption{Sum-rate comparison between IA beam selection with SD-based channel estimation and the one with OMP-based channel estimation.} \label{FIG3}
\end{figure}

\begin{figure}[tp]
\begin{center}
\vspace*{+1mm}\includegraphics[width=1\linewidth]{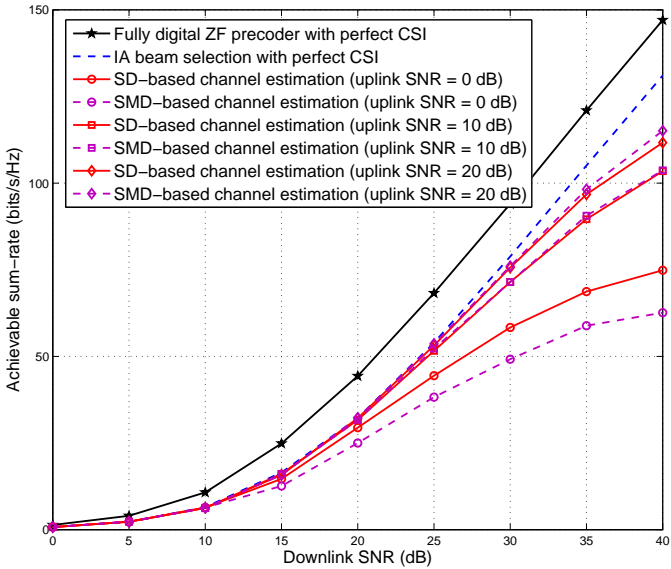}
\end{center}
\vspace*{-3mm}\caption{Sum-rate comparison between IA beam selection with SD-based channel estimation and the one with SMD-based channel estimation.} \label{FIG3}
\end{figure}

Next, we will evaluate the impact of different beamspace channel estimation schemes on beam selection. Here, we adopt the interference-aware (IA) beam selection scheme proposed in our previous work~\cite{gao16bs},  as it can guarantee that different users select different beams and support scenarios where ${{N_{{\rm{RF}}}} = K}$. Its key idea is to classify all users  into two user groups according to the potential inter-beam interferences. Specifically, if the beam with the largest power of a user is different from those of other users, this user will be regarded as a non-interference user (NIU). Otherwise, this user will be considered as an interference user (IU). For NIUs, the beams with largest power are directly selected, just like the traditional magnitude maximization beam selection scheme~\cite{sayeed2013beamspace}. For IUs, a low-complexity incremental algorithm is employed. It selects the beams one by one in an incremental order, and in each step the beam with the greatest contribution to the achievable sum-rate is selected.

Fig. 8 provides the sum-rate comparison between IA beam selection with SD-based channel estimation and the one with OMP-based channel estimation. Note that the IA beam selection scheme can select the beams with low interference based on the beamspace channel estimated by any scheme, such as the SD-based, OMP-based, and SMD-based channel estimation. We can observe that by utilizing the proposed SD-based channel estimation instead of the OMP-based one, IA beam selection can achieve better performance, especially when the uplink SNR is low. Moreover, when the uplink SNR is moderate (e.g., 10 dB), IA beam selection with SD-based channel estimation can achieve the sum-rate performance close to the one with perfect channel state information (CSI). Finally, it is worth pointing out that when the downlink SNR is sufficiently high (e.g., 40 dB), the sum-rate performance of IA beam selection with all channel estimation schemes will saturate. This is due to the fact that with high downlink SNR, the sum-rate performance is dominated by the channel estimation error instead of noise.

Fig. 9 provides the sum-rate comparison between IA beam selection with SD-based channel estimation and the one with SMD-based channel estimation~\cite{Hogan16}. From Fig. 9, we observe that when the uplink SNR is high (e.g., 20 dB), IA beam selection with SMD-based channel estimation can achieve higher sum-rate than the one with SD-based channel estimation. This is due to the fact that in SMD-based channel estimation, by scanning all the beams with low noise power, the weak beams can be also accurately estimated to efficiently eliminate interferences in IA beam selection. However, when the uplink SNR is not high (e.g., 0 dB), IA beam selection with SD-based channel estimation can slightly outperform the one with SMD-based channel estimation. This can be explained by the fact that in SMD-based channel estimation, the ${N}$ pilot symbols are respectively used to scan all ${N}$ beams without further process to suppress the noise, which makes the selected strong beams suboptimal in low uplink SNR region. This further verifies that the proposed SD-based channel estimation scheme can not only reduce the number of pilot symbols, but also achieve satisfying performance, especially when the uplink SNR is low.

%More importantly, when the uplink SNR is moderate (e.g., 10 dB), IA beam selection with SD-based channel estimation, which only requires 16 RF chains, can achieve the sum-rate performance not far away from the fully digital ZF precoder using 256 RF chains with perfect CSI. This means that mmWave massive MIMO systems with lens antenna array can achieve the near-optimal performance with significantly reduced number of RF chains and low pilot overhead in practice.

%Finally, we extend the proposed SD-based channel estimation to the scenarios where users employ multiple antennas. Fig. 9 shows the NMSE performance comparison, where we assume that each user has ${T = 4}$ antennas. The number of BS antennas, users, and instants for pilot transmission are still ${N=256}$, ${K=16}$, and ${Q = 96}$, respectively. The BS is assumed to employ ${{N_{{\rm{RF}}}} = KT = 64}$ RF chains. From Fig. 9, we can observe that the proposed SD-based channel estimation can still perform well, especially when the uplink SNR is low.

\section{Conclusions}\label{S6}
This paper investigates the channel estimation problem for  mmWave massive MIMO systems with lens antenna array. Specifically,  we first design an adaptive selecting network for mmWave massive MIMO systems with lens antenna array to formulate the beamspace channel estimation problem as a sparse signal recovery problem. Then, by utilizing the special structural characteristics of mmWave beamspace channel, we propose a SD-based channel estimation scheme with low pilot overhead. The performance analysis proves that the proposed SD-based channel estimation scheme can detect the support of sparse beamspace channel with higher accuracy than the classical CS algorithms. The  complexity analysis further shows that SD-based channel estimation  enjoys low complexity, which is comparable with that of LS algorithm. Simulation results verify that the proposed SD-based channel estimation scheme can achieve much better NMSE performance than conventional schemes, even in the low SNR region. This makes it more attractive for mmWave massive MIMO systems with lens antenna array. In our future work, we will extend the proposed SD-based channel estimation scheme to the scenarios where users employ multiple antennas.

\section*{Appendix A\\ Proof of Lemma 3}
To prove \textbf{Lemma 3}, we need to prove the the following \textbf{Lemma 4} at first.

\vspace*{+2mm} \noindent\textbf{Lemma 4}. {\it  Assume that the uplink noise vector ${{{\bf{\bar n}}_k}}$ in~(\ref{eq15}) follows the distribution ${{\cal C}{\cal N}\left( {0,\sigma _{{\rm{UL}}}^2{\bf{I}}} \right)}$, and define  ${\delta  \buildrel \Delta \over = \sqrt {2\sigma _{{\rm{UL}}}^2\left( {1 + \alpha } \right)\ln N} }$, where ${\alpha  > 0}$ is a constant. Then, we have}
\begin{equation}\label{eq51}
\Pr \left\{ {\mathop {\max }\limits_{1 \le n \le N} \left| {{\bf{\bar w}}_n^H{{{\bf{\bar n}}}_k}} \right| < \delta } \right\} \ge {\left( {1 - \frac{1}{{{N^\alpha }\sqrt {\pi \left( {1 + \alpha } \right)\ln N} }}} \right)^N},
\end{equation}
{\it where ${{{\bf{\bar w}}_n}}$ is the ${n}$th column of the combiner ${{\bf{\bar W}}}$ in~(\ref{eq15}).}

\vspace*{+2mm}
\textit{Proof:} Since ${Q < N}$, the ${N}$ columns ${\left\{ {{{{\bf{\bar w}}}_n}} \right\}_{n = 1}^N}$ of ${{\bf{\bar W}}}$ are correlated. Therefore, we can conclude that ${\left\{ {{\bf{\bar w}}_n^H{{{\bf{\bar n}}}_k}} \right\}_{n = 1}^N}$ are jointly Gaussian, leading to the following result
\begin{equation}\label{eq52}
\Pr \left\{ {\mathop {\max }\limits_{1 \le n \le N} \left| {{\bf{\bar w}}_n^H{{{\bf{\bar n}}}_k}} \right| < \delta } \right\} \ge \prod\limits_{n = 1}^N {\Pr \left\{ {\left| {{\bf{\bar w}}_n^H{{{\bf{\bar n}}}_k}} \right| < \delta } \right\}}.
\end{equation}
Since ${\left\{ {\left\| {{{{\bf{\bar w}}}_n}} \right\|_2^2} \right\}_{n = 1}^N = 1}$, each random Gaussian variable ${{\bf{\bar w}}_n^H{{\bf{\bar n}}_k}}$ has mean zero and variance ${\sigma _{{\rm{UL}}}^2}$. Then, we have
\begin{align}\label{eq53}
\Pr \left\{ {\left| {{\bf{\bar w}}_n^H{{{\bf{\bar n}}}_k}} \right| < \delta } \right\} = & 1 - \sqrt {\frac{2}{{\pi \sigma _{{\rm{UL}}}^2}}} \int_\delta ^{ + \infty } {{e^{ - \frac{{{x^2}}}{{2\sigma _{{\rm{UL}}}^2}}}}dx} \\ \nonumber
\mathop  \ge \limits^{\left( a \right)} & 1 - \sqrt {\frac{2}{\pi }} \frac{{{\sigma _{{\rm{UL}}}}}}{\delta }{e^{ - \frac{{{\delta ^2}}}{{2\sigma _{{\rm{UL}}}^2}}}}\\ \nonumber
\mathop  = \limits^{\left( b \right)} & 1 - \frac{1}{{{N^{\alpha  + 1}}\sqrt {\pi \left( {1 + \alpha } \right)\ln N} }},
\end{align}
where ${\left( a \right)}$ is true due to the fact that~\cite{ben2010coherence}
\begin{equation}\label{eq54}
\frac{1}{{\sqrt {2\pi } }}\int_y^{ + \infty } {{e^{ - \frac{{{x^2}}}{2}}}dx}  \le \frac{1}{{\sqrt {2\pi } y}}{e^{ - \frac{{{y^2}}}{2}}},
\end{equation}
and ${\left( b \right)}$ is true because ${\delta  \buildrel \Delta \over = \sqrt {2\sigma _{{\rm{UL}}}^2\left( {1 + \alpha } \right)\ln N} }$. Note that ${\alpha  > 0}$ and ${N}$ is the number of antennas. Therefore, it can be guaranteed that ${1/\left( {{N^{\alpha  + 1}}\sqrt {\pi \left( {1 + \alpha } \right)\ln N} } \right) < 1}$. Then, substituting~(\ref{eq53}) into~(\ref{eq52}), we can obtain the conclusion~(\ref{eq51}) of \textbf{Lemma 4}. \qed

By utilizing \textbf{Lemma 4}, we now can prove the conclusions in \textbf{Lemma 3}. Specifically, when the position of the strongest element ${{{{\tilde h}_{k,{n^ * }}}}}$ is correctly estimated, we have
\begin{equation}\label{eq56}
\left| {{\bf{\bar w}}_{{n^ * }}^H{{{\bf{\bar z}}}_k}} \right| > \mathop {\mathop {\max }\limits_{1 \le n \le N} }\limits_{n \ne {n^ * }} \left| {{\bf{\bar w}}_n^H{{{\bf{\bar z}}}_k}} \right|.
\end{equation}
According to~(\ref{eq15}), the left side of~(\ref{eq56}) can be lower-bounded by
\begin{align}\label{eq57}
\left| {{\bf{\bar w}}_{{n^ * }}^H{{{\bf{\bar z}}}_k}} \right| = & \left| {{{\tilde h}_{k,{n^ * }}} + \sum\limits_{n \ne {n^ * }} {{\bf{\bar w}}_{{n^ * }}^H{{\bf{\bar w}}_n}{{\tilde h}_{k,n}} + {\bf{\bar w}}_{{n^ * }}^H{{{\bf{\bar n}}}_k}} } \right|\\ \nonumber
\mathop  \ge \limits^{\left( a \right)} & \left| {{{\tilde h}_{k,{n^ * }}}} \right| - \left| {\sum\limits_{n \ne {n^ * }} {{\bf{\bar w}}_{{n^ * }}^H{{\bf{\bar w}}_n}{{\tilde h}_{k,n}} + {\bf{\bar w}}_{{n^ * }}^H{{{\bf{\bar n}}}_k}} } \right|\\ \nonumber
\mathop  \ge \limits^{\left( b \right)} & \left| {{{\tilde h}_{k,{n^ * }}}} \right| - \delta  - \left| {\sum\limits_{n \ne {n^ * }} {{\bf{\bar w}}_{{n^ * }}^H{{\bf{\bar w}}_n}{{\tilde h}_{k,n}}} } \right|\\ \nonumber
 \ge & \left| {{{\tilde h}_{k,{n^ * }}}} \right| - \delta  - \mu \sum\limits_{n \ne {n^ * }} {\left| {{{\tilde h}_{k,n}}} \right|} \\ \nonumber
\mathop  \ge \limits^{\left( c \right)} & \left| {{{\tilde h}_{k,{n^ * }}}} \right| - \delta  - \mu \eta \left| {{{\tilde h}_{k,{n^ * }}}} \right|,
\end{align}
where ${\left( a \right)}$ is true due to the triangle inequality, ${\left( b \right)}$ is valid because of the assumption ${\mathop {\max }\limits_{1 \le n \le N} \left| {{\bf{\bar w}}_n^H{{{\bf{\bar n}}}_k}} \right| < \delta }$, and ${\left( c \right)}$ is obtained by utilizing the structural characteristic of mmWave beamspace channel as proved in \textbf{Lemma 2}, where we have
\begin{align}\label{eq58}
\sum\limits_{n \ne {n^ * }} {\left| {{{\tilde h}_{k,n}}} \right|}  \le & \frac{{\sum\limits_{n = 1}^N {\left| {\frac{1}{{\sin \left( {\left( {2n - 1} \right)\pi /2N} \right)}}} \right|}  - \left| {\frac{1}{{\sin \left( {\pi /2N} \right)}}} \right|}}{{\left| {\frac{1}{{\sin \left( {\pi /2N} \right)}}} \right|}}\left| {{{\tilde h}_{k,{n^ * }}}} \right| \\ \nonumber
 = & \eta \left| {{{\tilde h}_{k,{n^ * }}}} \right|.
\end{align}

On the other hand, the right side of~(\ref{eq56}) can be upper-bounded by
\begin{align}\label{eq59}
& \mathop {\mathop {\max }\limits_{1 \le n \le N} }\limits_{n \ne {n^ * }} \left| {{\bf{\bar w}}_n^H{{{\bf{\bar z}}}_k}} \right| \! = \! \mathop {\mathop {\max }\limits_{1 \le n \le N} }\limits_{n \ne {n^ * }} \left| {{\bf{\bar w}}_n^H{{{\bf{\bar n}}}_k} \! + \! \sum\limits_{j = 1}^N {{\bf{\bar w}}_n^H{{{\bf{\bar w}}}_j}{{\tilde h}_{k,j}}} } \right|\\ \nonumber
\mathop  \le \limits^{\left( a \right)} & \delta  + \left| {{{\tilde h}_{k,{n^{ *  * }}}}} \right| + \mathop {\mathop {\max }\limits_{1 \le n \le N} }\limits_{n \ne {n^ * }} \left| {\sum\limits_{j = 1,j \ne {n^{ *  * }}}^N {{\bf{\bar w}}_n^H{{{\bf{\bar w}}}_j}{{\tilde h}_{k,j}}} } \right|\\ \nonumber
 \le & \delta  + \left| {{{\tilde h}_{k,{n^{ *  * }}}}} \right| + \mu \left| {{{\tilde h}_{k,{n^ * }}}} \right| + \mu \eta \left| {{{\tilde h}_{k,{n^ * }}}} \right| - \mu \left| {{{\tilde h}_{k,{n^{ *  * }}}}} \right|\\ \nonumber
\mathop  \le \limits^{\left( b \right)} & \delta  + \mu \left| {{{\tilde h}_{k,{n^ * }}}} \right| + \mu \eta \left| {{{\tilde h}_{k,{n^ * }}}} \right| + \left( {1 - \mu } \right)\kappa \left| {{{\tilde h}_{k,{n^ * }}}} \right|,
\end{align}
where ${{{{\tilde h}_{k,{n^{ *  * }}}}}}$ is the second strongest element of ${{{\bf{\tilde h}}_k}}$, ${\left( a \right)}$ is true due to the fact that ${n}$ may be equal to ${{n^{ *  * }}}$, leading to ${{\bf{\bar w}}_{{n^{ *  * }}}^H{{\bf{\bar w}}_{{n^{ *  * }}}} = 1}$, and ${\left( b \right)}$ can be also obtained from \textbf{Lemma 2}, where we have
\begin{equation}\label{eq60}
\left| {{{\tilde h}_{k,{n^{ *  * }}}}} \right| \le \frac{{\sin \left( {\pi /2N} \right)}}{{\sin \left( {3\pi /2N} \right)}}\left| {{{\tilde h}_{k,{n^ * }}}} \right| = \kappa \left| {{{\tilde h}_{k,{n^ * }}}} \right|.
\end{equation}
Combine~(\ref{eq57}) and~(\ref{eq59}) will yield
\begin{equation}\label{eq61}
\left| {{{\tilde h}_{k,{n^ * }}}} \right| \ge \frac{{2\delta }}{{\left( {1 - \mu } \right)\left( {1 - \kappa } \right) - 2\mu \eta }}.
\end{equation}
Then, by utilizing \textbf{Lemma 4}, we can conclude that when
\begin{equation}\label{eq62}
\left| {{{\tilde h}_{k,{n^ * }}}} \right| \ge \frac{{\sqrt {8\sigma _{{\rm{UL}}}^2\left( {1 + \alpha } \right)\ln N} }}{{\left( {1 - \mu } \right)\left( {1 - \kappa } \right) - 2\mu \eta }},
\end{equation}
the probability that the position of the strongest element is correctly estimated is lower-bounded by~(\ref{eq28}). \qed

Finally, it is worth pointing out that the conclusions in \textbf{Lemma 3} can be also extended to the scenario with NLoS components, where the only changes are the expressions of ${\eta }$ and ${\kappa }$. This is due to fact that all channel components of ${{{\mathbf{\tilde{h}}}_{k}}}$ are approximately orthogonal to each other as proved in \textbf{Lemma 1}. Therefore we can always find the upper-bounds of ${\sum\limits_{n \ne {n^ * }} {\left| {{{\tilde h}_{k,n}}} \right|} /\left| {{{\tilde h}_{k,{n^ * }}}} \right|}$ and ${\left| {{{\tilde h}_{k,{n^{ *  * }}}}} \right|/\left| {{{\tilde h}_{k,{n^ * }}}} \right|}$ by utilizing the structural characteristic of mmWave beamspace channel as proved in \textbf{Lemma 2}.

\bibliography{IEEEabrv,Gao1Ref}

\begin{thebibliography}{10}
\providecommand{\url}[1]{#1}
\csname url@samestyle\endcsname
\providecommand{\newblock}{\relax}
\providecommand{\bibinfo}[2]{#2}
\providecommand{\BIBentrySTDinterwordspacing}{\spaceskip=0pt\relax}
\providecommand{\BIBentryALTinterwordstretchfactor}{4}
\providecommand{\BIBentryALTinterwordspacing}{\spaceskip=\fontdimen2\font plus
\BIBentryALTinterwordstretchfactor\fontdimen3\font minus
  \fontdimen4\font\relax}
\providecommand{\BIBforeignlanguage}[2]{{%
\expandafter\ifx\csname l@#1\endcsname\relax
\typeout{** WARNING: IEEEtran.bst: No hyphenation pattern has been}%
\typeout{** loaded for the language `#1'. Using the pattern for}%
\typeout{** the default language instead.}%
\else
\language=\csname l@#1\endcsname
\fi
#2}}
\providecommand{\BIBdecl}{\relax}
\BIBdecl

\bibitem{han2015large}
S.~Han, C.-L. I, Z.~Xu, and C.~Rowell, ``Large-scale antenna systems with
  hybrid precoding analog and digital beamforming for millimeter wave 5{G},''
  \emph{{IEEE} Commun. Mag.}, vol.~53, no.~1, pp. 186--194, Jan. 2015.

\bibitem{pi2011introduction}
Z.~Pi and F.~Khan, ``An introduction to millimeter-wave mobile broadband
  systems,'' \emph{{IEEE} Commun. Mag.}, vol.~49, no.~6, pp. 101--107, Jun.
  2011.

\bibitem{rusek13}
F.~Rusek, D.~Persson, B.~K. Lau, E.~G. Larsson, T.~L. Marzetta, O.~Edfors, and
  F.~Tufvesson, ``Scaling up {MIMO}: Opportunities and challenges with very
  large arrays,'' \emph{{IEEE} Signal Process. Mag.}, vol.~30, no.~1, pp.
  40--60, Jan. 2013.

\bibitem{wei2014key}
L.~Wei, R.~Q. Hu, Y.~Qian, and G.~Wu, ``Key elements to enable millimeter wave
  communications for 5{G} wireless systems,'' \emph{IEEE Wireless Commun.},
  vol.~21, no.~6, pp. 136--143, Dec. 2014.

\bibitem{alkhateeb2014mimo}
A.~Alkhateeb, J.~Mo, N.~Gonz{\'a}lez-Prelcic, and R.~W. Heath, ``{MIMO}
  precoding and combining solutions for millimeter-wave systems,'' \emph{{IEEE}
  Commun. Mag.}, vol.~52, no.~12, pp. 122--131, Dec. 2014.

\bibitem{brady2013beamspace}
J.~Brady, N.~Behdad, and A.~Sayeed, ``Beamspace {MIMO} for millimeter-wave
  communications: System architecture, modeling, analysis, and measurements,''
  \emph{IEEE Trans. Ant. and Propag.}, vol.~61, no.~7, pp. 3814--3827, Jul.
  2013.

\bibitem{zeng16mmwave}
Y.~Zeng and R.~Zhang, ``Millimeter wave {MIMO} with lens antenna array: {A} new
  path division multiplexing paradigm,'' \emph{{IEEE} Trans. Commun.}, vol.~64,
  no.~4, pp. 1557--1571, Apr. 2016.

\bibitem{Zeng_2014}
Y.~Zeng, R.~Zhang, and Z.~N. Chen, ``Electromagnetic lens-focusing antenna
  enabled massive {MIMO}: Performance improvement and cost reduction,''
  \emph{{IEEE} J. Sel. Areas Commun.}, vol.~32, no.~6, pp. 1194--1206, Jun.
  2014.

\bibitem{Behdad10}
N.~Behdad and A.~Sayeed, ``Continuous aperture phased {MIMO}: {B}asic theory
  and applications,'' in \emph{Proc. Allerton Conference}, Sep. 2010, pp.
  1196--1203.

\bibitem{sayeed2013beamspace}
A.~Sayeed and J.~Brady, ``Beamspace {MIMO} for high-dimensional multiuser
  communication at millimeter-wave frequencies,'' in \emph{Proc. {IEEE}
  GLOBECOM}, Dec. 2013, pp. 3679--3684.

\bibitem{amadorilow}
P.~Amadori and C.~Masouros, ``Low {RF}-complexity millimeter-wave
  beamspace-{MIMO} systems by beam selection,'' \emph{{IEEE} Trans. Commun.},
  vol.~63, no.~6, pp. 2212--2222, Jun. 2015.

\bibitem{gao16bs}
X.~Gao, L.~Dai, Z.~Chen, Z.~Wang, and Z.~Zhang, ``Near-optimal beam selection
  for beamspace mm{W}ave massive {MIMO} systems,'' \emph{{IEEE} Commun. Lett.},
  vol.~20, no.~5, pp. 1054--1057, May 2016.

\bibitem{alkhateeb2014channel}
A.~Alkhateeb, O.~El~Ayach, G.~Leus, and R.~W. Heath, ``Channel estimation and
  hybrid precoding for millimeter wave cellular systems,'' \emph{\it IEEE J.
  Sel. Top. Signal Process.}, vol.~8, no.~5, pp. 831--846, Oct. 2014.

\bibitem{alkhateeb2015compressed}
A.~Alkhateeb, G.~Leus, and R.~W. Heath, ``Compressed sensing based multi-user
  millimeter wave systems: How many measurements are needed?'' in \emph{Proc.
  IEEE {ICASSP}}, Apr. 2015, pp. 2909--2913.

\bibitem{love15}
T.~Kim and D.~J. Love, ``Virtual {A}o{A} and {A}o{D} estimation for sparse
  millimeter wave {MIMO} channels,'' in \emph{Proc. IEEE {SPAWC} Workshops},
  Jun. 2015, pp. 146--150.

\bibitem{gao2015mmwave}
Z.~Gao, L.~Dai, D.~Mi, Z.~Wang, M.~A. Imran, and M.~Z. Shakir, ``Mmwave
  massive-{MIMO}-based wireless backhaul for the 5{G} ultra-dense network,''
  \emph{IEEE Wireless Commun.}, vol.~22, no.~5, pp. 13--21, Oct. 2015.

\bibitem{bajwa2010compressed}
W.~U. Bajwa, J.~Haupt, A.~Sayeed, and R.~Nowak, ``Compressed channel sensing:
  {A} new approach to estimating sparse multipath channels,'' \emph{Proc.
  IEEE}, vol.~98, no.~6, pp. 1058--1076, Jun. 2010.

\bibitem{el2013spatially}
O.~El~Ayach, S.~Rajagopal, S.~Abu-Surra, Z.~Pi, and R.~W. Heath, ``Spatially
  sparse precoding in millimeter wave {MIMO} systems,'' \emph{{IEEE} Trans.
  Wireless Commun.}, vol.~13, no.~3, pp. 1499--1513, Mar. 2014.

\bibitem{gao16turbo}
X.~Gao, L.~Dai, C.~Yuen, and Z.~Wang, ``Turbo-like beamforming based on tabu
  search algorithm for millimeter-wave massive {MIMO} systems,'' \emph{{IEEE}
  Trans. Veh. Technol.}, vol.~65, no.~7, pp. 5731--5737, Jul. 2016.

\bibitem{gao15energy}
X.~Gao, L.~Dai, S.~Han, C.-L. I, and R.~W. Heath, ``Energy-efficient hybrid
  analog and digital precoding for mm{W}ave {MIMO} systems with large antenna
  arrays,'' \emph{{IEEE} J. Sel. Areas Commun.}, vol.~34, no.~4, pp. 998--1009,
  Apr. 2016.

\bibitem{Hogan16}
J.~Hogan and A.~Sayeed, ``Beam selection for performance-complexity
  optimization in high-dimension {MIMO} systems,'' in \emph{Proc. CISS}, Mar.
  2016, pp. 337--342.

\bibitem{Yang16}
L.~Yang, Y.~Zeng, and R.~Zhang, ``Efficient channel estimation for millimeter
  wave {MIMO} with limited {RF} chains,'' in \emph{Proc. IEEE ICC}, May 2016,
  pp. 1--6.

\bibitem{song2013beamspace}
G.~H. Song, J.~Brady, and A.~Sayeed, ``Beamspace {MIMO} transceivers for
  low-complexity and near-optimal communication at mm-wave frequencies,'' in
  \emph{Proc. IEEE {ICASSP}}, Apr. 2013, pp. 4394--4398.

\bibitem{tropp2006algorithms1}
J.~A. Tropp, A.~C. Gilbert, and M.~J. Strauss, ``Algorithms for simultaneous
  sparse approximation. {P}art {I}: Greedy pursuit,'' \emph{Signal Processing},
  vol.~86, no.~3, pp. 572--588, Mar. 2006.

\bibitem{gao16iccc}
L.~Dai, X.~Gao, S.~Han, C.-L. I, and X.~Wang, ``Beamspace channel estimation
  for millimeter-wave massive {MIMO} systems with lens antenna array,'' in
  \emph{Proc. IEEE/CIC ICCC}, Jul. 2016, pp. 1--6.

\bibitem{rappaport2011state}
T.~S. Rappaport, J.~N. Murdock, and F.~Gutierrez, ``State of the art in
  60-{GH}z integrated circuits and systems for wireless communications,''
  \emph{Proc. IEEE}, vol.~99, no.~8, pp. 1390--1436, Aug. 2011.

\bibitem{brady2014prototype}
J.~Brady, P.~Thomas, D.~Virgilio, and A.~Sayeed, ``Beamspace {MIMO} prototype
  for low-complexity {G}igabit/s wireless communication,'' in \emph{Proc. IEEE
  {SPAWC} Workshops}, Jun. 2014, pp. 135--139.

\bibitem{brady2016multi}
J.~Brady, J.~Hogan, and A.~Sayeed, ``Multi-beam {MIMO} prototype for real-time
  multiuser communication at 28 {GH}z,'' in \emph{Proc. IEEE {GLOBECOM}
  Workshops}, Dec. 2016, pp. 1--6.

\bibitem{tse2005fundamentals}
D.~Tse and P.~Viswanath, \emph{Fundamentals of wireless communication}.\hskip
  1em plus 0.5em minus 0.4em\relax Cambridge University Press, 2005.

\bibitem{Kotecha_2004}
J.~Kotecha and A.~Sayeed, ``Transmit signal design for optimal estimation of
  correlated {MIMO} channels,'' \emph{{IEEE} Trans. Signal Process.}, vol.~52,
  no.~2, pp. 546--557, Feb. 2004.

\bibitem{Dongwoon12}
B.~Dongwoon, P.~Cheolhee, L.~Jungwon, N.~Hoang, J.~Singh, A.~Gupta, P.~Zhouyue,
  K.~Taeyoon, L.~Chaiman, K.~Min-Goo, and I.~Inyup, ``{LTE}-{A}dvanced modem
  design: {C}hallenges and perspectives,'' \emph{{IEEE} Commun. Mag.}, vol.~50,
  no.~2, pp. 178--186, Feb. 2012.

\bibitem{xie2016unified}
H.~Xie, F.~Gao, S.~Zhang, and S.~Jin, ``A unified transmission strategy for
  {TDD}/{FDD} massive {MIMO} systems with spatial basis expansion model,''
  \emph{{IEEE} Trans. Veh. Technol.}, vol.~66, no.~4, pp. 3170--3184, Apr.
  2017.

\bibitem{Tucker_phase}
N.~Tucker, ``A low cost electro-mechanical phase shifter design, including a
  brief summary of solid state methods,'' available at www.activefrance.com.

\bibitem{Liang_2014}
L.~Liang, W.~Xu, and X.~Dong, ``Low-complexity hybrid precoding in massive
  multiuser {MIMO} systems,'' \emph{{IEEE} Wireless Commun. Lett.}, vol.~3,
  no.~6, pp. 653--656, Oct. 2014.

\bibitem{mendez2016hybrid}
R.~M{\'e}ndez-Rial, C.~Rusu, N.~Gonz{\'a}lez-Prelcic, A.~Alkhateeb, and R.~W.
  Heath, ``Hybrid {MIMO} architectures for millimeter wave communications:
  {P}hase shifters or switches?'' \emph{IEEE Access}, vol.~4, pp. 247--267,
  Jan. 2016.

\bibitem{tropp2007signal}
J.~A. Tropp and A.~C. Gilbert, ``Signal recovery from random measurements via
  orthogonal matching pursuit,'' \emph{{IEEE} Trans. Inf. Theory}, vol.~53,
  no.~12, pp. 4655--4666, Dec. 2007.

\bibitem{zhou2016channel}
Z.~Zhou, J.~Fang, L.~Yang, H.~Li, Z.~Chen, and S.~Li, ``Channel estimation for
  millimeter-wave multiuser {MIMO} systems via {PARAFAC} decomposition,''
  \emph{{IEEE} Trans. Wireless Commun.}, vol.~15, no.~11, pp. 7501--7516, Nov.
  2016.

\bibitem{gao16broadband}
Z.~Gao, L.~Dai, C.~Hu, and Z.~Wang, ``Channel estimation for millimeter-wave
  massive {MIMO} with hybrid precoding over frequency-selective fading
  channels,'' \emph{{IEEE} Commun. Lett.}, vol.~20, no.~6, pp. 1259--1262, Jun.
  2016.

\bibitem{xie2016overview}
H.~Xie, F.~Gao, and S.~Jin, ``An overview of low-rank channel estimation for
  massive {MIMO} systems,'' \emph{IEEE Access}, vol.~4, pp. 7313--7321, Nov.
  2016.

\bibitem{Sayeed_2002}
A.~Sayeed, ``Deconstructing multiantenna fading channels,'' \emph{{IEEE} Trans.
  Signal Process.}, vol.~50, no.~10, pp. 2563--2579, Oct. 2002.

\bibitem{mendez2015channel}
R.~Mendez-Rial, C.~Rusu, A.~Alkhateeb, N.~Gonz{\'a}lez-Prelcic, and R.~W.
  Heath, ``Channel estimation and hybrid combining for mm{W}ave: {P}hase
  shifters or switches?'' in \emph{Proc. IEEE {ITA} Workshops}, Feb. 2015, pp.
  90--97.

\bibitem{ben2010coherence}
Z.~Ben-Haim, Y.~C. Eldar, and M.~Elad, ``Coherence-based performance guarantees
  for estimating a sparse vector under random noise,'' \emph{{IEEE} Trans.
  Signal Process.}, vol.~58, no.~10, pp. 5030--5043, Oct. 2010.

\end{thebibliography}

\vspace*{-10mm}
\begin{IEEEbiography}[{\includegraphics[width=1in,height=1.25in,clip,keepaspectratio]{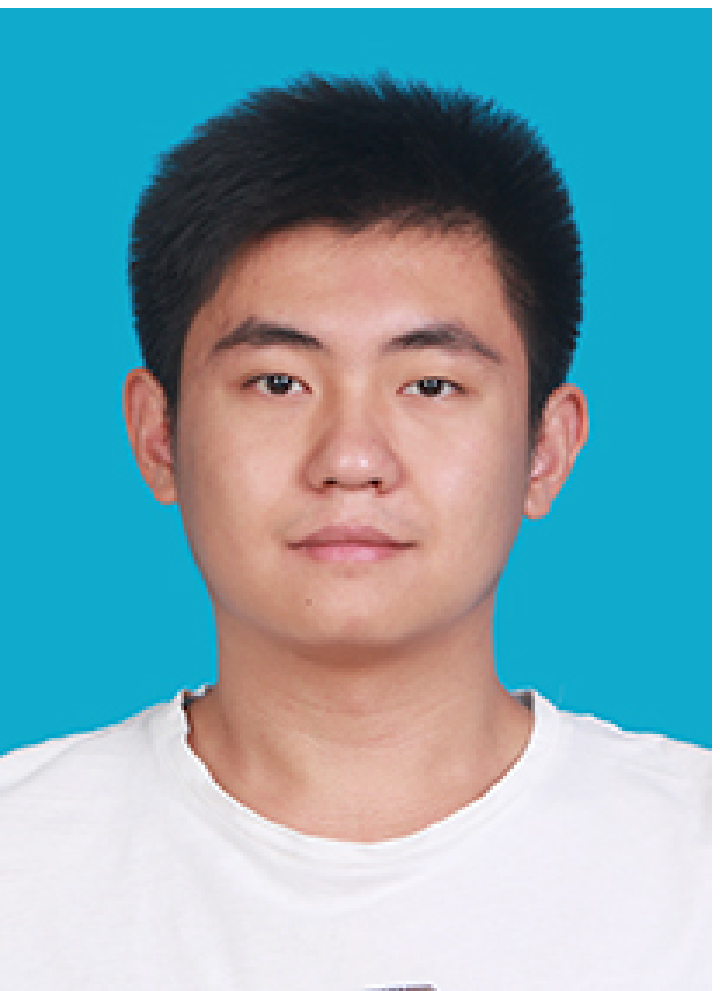}}]
{Xinyu Gao}(S'14) received the B.E. degree of Communication
Engineering from Harbin Institute of Technology, Heilongjiang, China in 2014. He is currently working towards Ph. D. degree in Electronic Engineering from Tsinghua University, Beijing, China. His research interests include massive MIMO and mmWave communications, with the emphasis on signal detection and precoding. He has published several journal and conference papers in IEEE Journal on Selected Areas in Communications, IEEE Transaction on Vehicular Technology, IEEE ICC, IEEE GLOBECOM, etc. He has won the national scholarship in 2015, the IEEE WCSP Best Paper Award in 2016.
\end{IEEEbiography}

\vspace*{-10mm}
\begin{IEEEbiography}[{\includegraphics[width=1in,height=1.25in,clip,keepaspectratio]{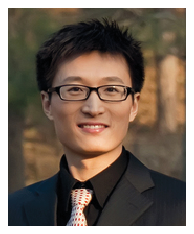}}]
{Linglong Dai} (M'11- SM'14) received the B.S. degree from Zhejiang University in 2003, the M.S. degree (with the highest honor) from the China Academy of Telecommunications Technology (CATT) in 2006, and the Ph.D. degree (with the highest honor) from Tsinghua University, Beijing, China, in 2011. From 2011 to 2013, he was a postdoctoral fellow with the Department of Electronic Engineering, Tsinghua University, where he has been an assistant professor since July 2013. His research interests are in wireless communications, with a focus on multi-carrier techniques, multi-antenna techniques, and multi-user techniques. He has published over 50 IEEE journal papers and over 30 IEEE conference papers. He has received the Outstanding Ph.D. Graduate of Tsinghua University Award in 2011, the Excellent Doctoral Dissertation of Beijing Award in 2012, the IEEE ICC Best Paper Award in 2013, the National Excellent Doctoral Dissertation Nomination Award in 2013, the IEEE ICC Best Paper Award in 2014, the URSI Young Scientists Award in 2014, the IEEE Transactions on Broadcasting Best Paper Award in 2015, the IEEE RADIO Young Scientists Award in 2015. He currently serves as a co-chair of the IEEE Special Interest Group (SIG) on Signal Processing Techniques in 5G Communication Systems.
\end{IEEEbiography}

\vspace*{-10mm}
\begin{IEEEbiography}[{\includegraphics[width=1in,height=1.25in,clip,keepaspectratio]{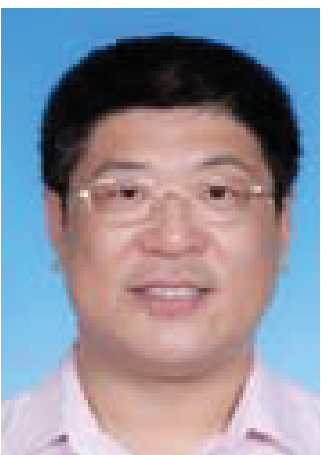}}]
{Shuangfeng Han} received his M.S. and Ph.D. degrees in electrical engineering from Tsinghua University in 2002 and 2006 respectively. He joined Samsung Electronics as a senior engineer in 2006 working on MIMO, MultiBS MIMO, etc. From 2012, he is a senior project manager in the Green Communication Research Center at the China Mobile Research Institute. His research interests are green 5G, massive MIMO, full duplex, NOMA and EE-SE co-design. Currently, he is the vice chair of wireless technology work group of China's IMT-2020 (5G) promotion group.
\end{IEEEbiography}

\vspace*{-10mm}
\begin{IEEEbiography}[{\includegraphics[width=1in,height=1.25in,clip,keepaspectratio]{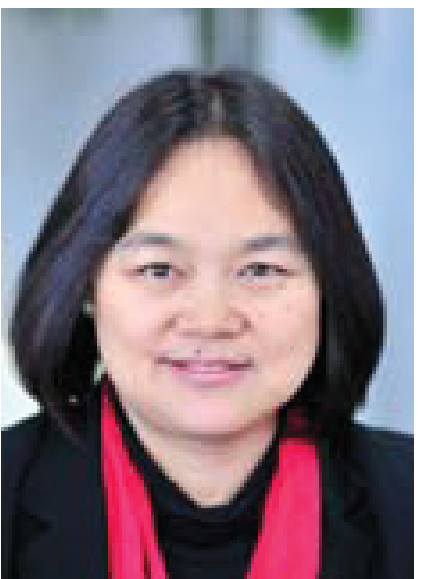}}]
{Chih-Lin I} received her Ph.D. degree in electrical engineering from Stanford University. She has been working at multiple world-class companies and research institutes leading the R${{\rm{\& }}}$D, including AT${{\rm{\& }}}$T Bell Labs; AT${{\rm{\& }}}$T HQ, ITRI of Taiwan, and ASTRI of Hong Kong. She received the IEEE Trans. on Commu. Stephen Rice Best Paper Award and is a winner of the CCCP National 1000 Talent program. Currently, she is China Mobile's chief scientist of wireless technologies and has established the Green Communications Research Center, spearheading major initiatives including key 5G technology R${{\rm{\& }}}$D; high energy efficiency system architectures, technologies and devices; green energy; and C-RAN and soft base stations. She was an elected Board Member of IEEE ComSoc, Chair of the ComSoc Meetings and Conferences Board, and Founding Chair of the IEEE WCNC Steering Committee. She is currently an Executive Board Member of GreenTouch and a Network Operator Council Member of ETSI NFV. Her research interests are green communications, C-RAN, network convergence, bandwidth refarming, EE-SE co-design, massive MIMO, and active antenna arrays.
\end{IEEEbiography}

\vspace*{-10mm}
\begin{IEEEbiography}[{\includegraphics[width=1in,height=1.25in,clip,keepaspectratio]{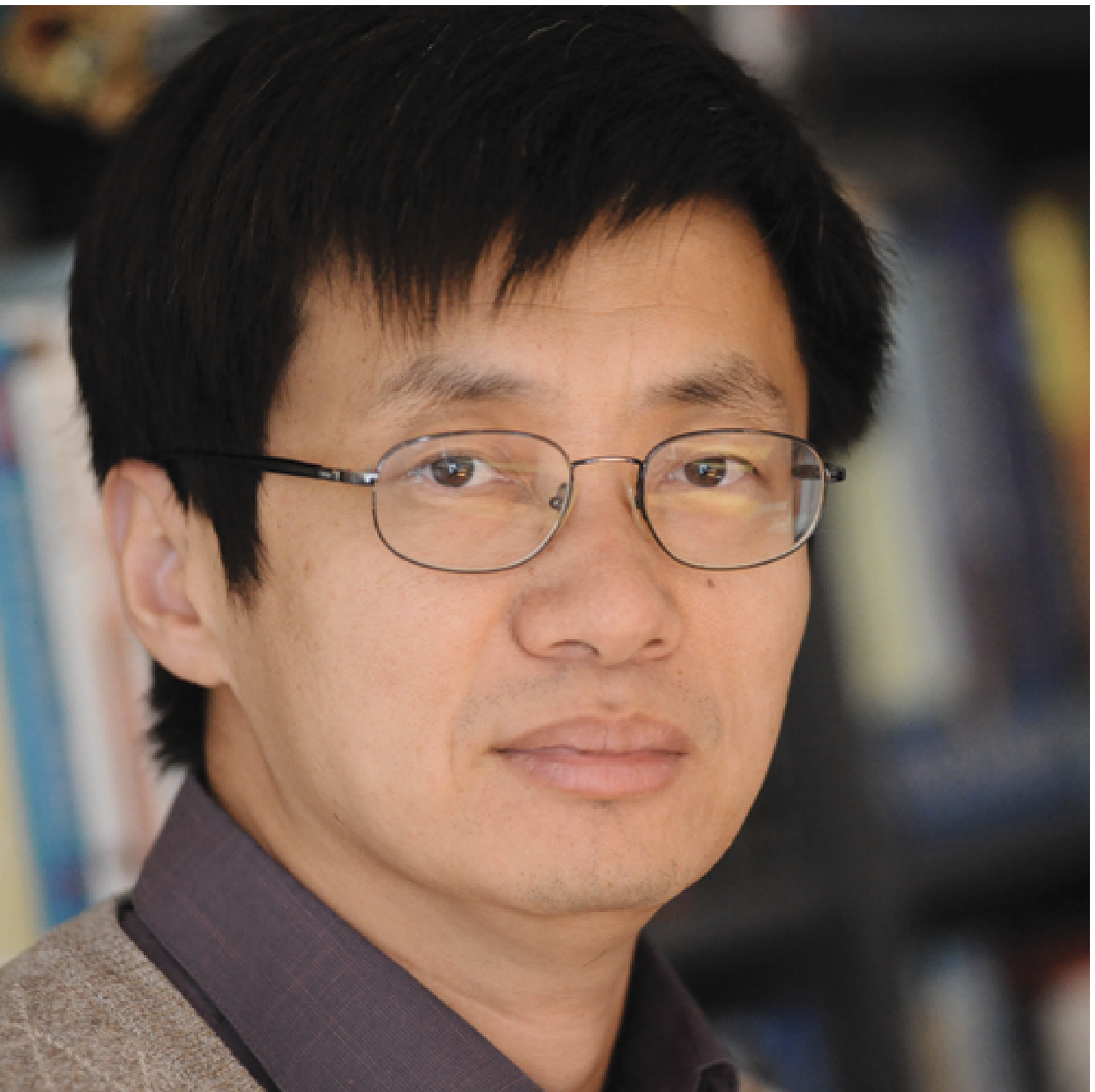}}]
{Xiaodong Wang} (S'98 - M'98 - SM'04 - F'08) received the Ph.D degree in Electrical Engineering from Princeton University. He is a Professor of  Electrical Engineering at Columbia University in New York. Dr. Wang's research interests fall in the general areas of computing, signal processing and communications, and has published extensively in these areas. Among his publications is a book entitled ``Wireless Communication Systems: Advanced Techniques for Signal Reception'', published by Prentice Hall in 2003.  His current research interests include wireless communications,  statistical signal processing, and genomic signal processing. Dr. Wang received the 1999 NSF CAREER Award, the 2001 IEEE Communications Society and Information Theory Society Joint Paper Award, and the 2011 IEEE Communication Society Award for Outstanding Paper on New Communication Topics. He has served  as an Associate Editor for the {\em IEEE Transactions on Communications}, the {\em IEEE Transactions on Wireless Communications}, the {\em IEEE Transactions on Signal Processing}, and   the {\em IEEE Transactions on Information Theory}. He is a Fellow of the IEEE and listed as an ISI Highly-cited Author.
\end{IEEEbiography}

\end{document}